\documentclass[12pt]{article}
\usepackage{amsmath}
\usepackage{graphicx}
\usepackage{natbib}
\usepackage{url} % not crucial - just used below for the URL 
\usepackage{booktabs}       % professional-quality tables
\usepackage{amsfonts}       % blackboard math symbols
\usepackage{nicefrac}       % compact symbols for 1/2, etc.
\usepackage{microtype}      % microtypography
\usepackage{algorithm}
\usepackage{algorithmic}
\usepackage{multirow}
\usepackage{subcaption}

\newcommand{\blind}{0}

\begin{document}

\def\spacingset#1{\renewcommand{\baselinestretch}%
{#1}\small\normalsize} \spacingset{1}

%%%%%%%%%%%%%%%%%%%%%%%%%%%%%%%%%%%%%%%%%%%%%%%%%%%%%%%%%%%%%%%%%%%%%%%%%%%%%%

\if0\blind
{
  \title{\bf ABC Learning of Hawkes Processes with Missing or Noisy Event Times}
  \author{Isabella Deutsch\thanks{
    The authors gratefully acknowledge \textit{please remember to list all relevant funding sources in the unblinded version}}\hspace{.2cm} \\
    School of Mathematics, University of Edinburgh\\
    and \\
    Gordon J. Ross \\
     School of Mathematics, University of Edinburgh}
  \maketitle
} \fi

\if1\blind
{
  \bigskip
  \bigskip
  \bigskip
  \begin{center}
    {\LARGE\bf ABC Learning of Hawkes Processes with Missing or Noisy Event Times}
\end{center}
  \medskip
} \fi

\bigskip
\begin{abstract}
 The self-exciting Hawkes process is widely used to model events which occur in bursts. However, many real world data sets contain missing events and/or noisily observed event times, which we refer to as data distortion. The presence of such distortion can severely bias the learning of the Hawkes process parameters. To circumvent this, we propose modeling the distortion function explicitly. This leads to a model with an intractable likelihood function which makes it difficult to deploy standard parameter estimation techniques. As such, we develop the ABC-Hawkes algorithm which is a novel approach to estimation based on Approximate Bayesian Computation (ABC) and Markov Chain Monte Carlo. This allows the parameters of the Hawkes process to be learned in settings where conventional methods induce substantial bias or are inapplicable. The proposed approach is shown to perform well on both real and simulated data.
\end{abstract}

\noindent%
\vfill

\newpage
\spacingset{1.5} % DON'T change the spacing!

\section{Introduction}

Hawkes processes \citep{hawkes_spectra_1971} are a class of point processes that are used to model event data when the events can occur in clusters or bursts. Typical examples include earthquakes  \citep{omi_estimating_2014}, stock transaction \citep{rambaldi_role_2017}, crime \citep{Shelton}, and social media \citep{lai_topic_2016, mei_neural_2017}. 

The Hawkes process is defined by a conditional intensity function $\lambda(\cdot)$ which controls the probability of events occurring at each interval in time, based on the previous history of the process. The conditional nature of the intensity function allows the intensity to increase for a short period of time whenever an event occurs, which results in a higher probability of further events occurring, hence creating event clusters and bursts.

Typically $\lambda(\cdot)$ is assigned a parametric form which allows for a relatively straightforward estimation using maximum likelihood or Bayesian methods \citep{Veen_2008_estimation,rasmussen_bayesian_2013}. Alternatively, a nonparametric estimation of the intensity function has been proposed for a variety of settings: stochastic intensity functions \citep{donnet_nonparametric_2018} or using kernel methods \citep{choi_nonparametric_1999}, LTSM neural networks \citep{mei_neural_2017}, and GANs \citep{xiao_wasserstein_2017}.

Most applications of the Hawkes process assume that the entire history of events has been accurately observed. However, many real world applications suffer from missing data where some events are undetected \citep{mei_imputing_2019}, or from noisy data where the recorded event times are inaccurate \citep{trouleau_learning_2019}. We refer to this as data distortion, and it can occur for several reasons. For example, in earthquake catalogues it is well known that the occurrence of large earthquakes has a masking effect which reduces the probability of subsequent earthquakes being detected for a period of time \citep{helmstetter_comparison_2006, omi_estimating_2014, arcangelis_overlap_2018}. Alternatively, event data may simply not be available for a certain interval of time, such as in a terrorism data set considered by \citet{tucker_handling_2019}.

Distorted (e.g. missing or noisy) data causes serious problems for Hawkes processes. If the model parameters are learned using only the observed data then the estimation of $\lambda(\cdot)$ may be severely biased. As such, a principled learning algorithm needs to consider the impact of the distortion.  So far, there has only been limited work on learning Hawkes processes in the presence of distorted data. The main exception to this is when the distortion takes the form of gaps in the observations where no events are detected at all  for a certain period \citep{Le, Shelton}. In this context, \citet{tucker_handling_2019} develop a Bayesian estimation algorithm which uses MCMC to impute missing events, and a similar approach is proposed by \cite{mei_imputing_2019} using particle smoothing. \citet{linderman_bayesian_2017} view the true generating process as a latent variable, which can be learned through sequential Monte Carlo techniques. Other examples look at specific instances of censored data \citep{xu_learning_2017} or asynchronous data \citep{upadhyay_deep_2018, trouleau_learning_2019}.% Although interesting, these approaches either mention  missing \textit{or} noisy data of a specific form and do not offer a unified framework for distorted data.

%\citet{mei_neural_2017} introduce ``neural Hawkes'', which uses neural networks to learn the intensity function of a point process. They ``hope'' that their approach readily applies to Hawkes process data with missing events by automatically adjusting to the data distortion. This is not obvious, as the distorted data from a Hawkes process cannot be modelled as Hawkes process with modified intensity due to the self-exciting nature of the data (see Section~\ref{sec_missing_data}). While \citet{mei_neural_2017} demonstrate their approach on one example, they do not explicitly model the data distorting mechanism.

In this paper, we present a more general approach for estimating Hawkes processes in the presence of distortion, which can handle a much wider class of distortion scenarios, including the case of gaps in the observed data, the case where there is a reduced probability of detecting events during some time period, and the case of noise in the recorded observation times. Our approach assumes the existence of a general distortion function $h(\cdot)$ which specifies the type of distortion that is present. The resulting Hawkes process likelihood is computationally intractable, since the self-excitation component involves triggering from the (unobserved) true event times, which must be integrated out to give the likelihood of the observed data. To solve this problem, we propose a novel estimation scheme using Approximate Bayesian Computation \citep[ABC,][]{marin_approximate_2012} to learn the Hawkes intensity in the presence of distortion. The resulting algorithm, ABC-Hawkes, is based on applying ABC using particular summary statistics of the Hawkes process, with separate convergence thresholds for each statistic. 

%, the resulting data model turns out to be computationally intractable: It is not even possible to evaluate the resulting likelihood function, due to the need to marginalize over the unknown missing events. Similarly, in the case of noisy data, $h(\cdot)$ adds error to each event. Events would nonetheless trigger according to their original time, not the observed one. Again, this leads to an intractable likelihood. This has limited the use of similar attempts to model distorted data, since there is no known method for learning the posterior of a Hawkes process when the likelihood function is intractable \citep{omi_estimating_2014}.

%This paper introduces ABC to Hawkes process modeling, where previous applications of this method are sparse. One rare example by \citet{ertekin_reactive_2015} uses ABC for Hawkes processes. However, the conceptual need for ABC in their setting is not evident to us as the likelihood can be evaluated, and hence exact MCMC algorithms could be used. Instead, our unique framework provides an innovative, new approach to handling distorted data for Hawkes processes where the likelihood cannot be evaluated. It enables us to sensibly estimate posterior densities where current methods induce substantial bias or are inapplicable altogether.

The paper is organized as follows. In Section~\ref{sec_problemoverview} we introduce the Hawkes process and the distorted data setting. Section~\ref{sec_ABC} summarizes ABC and introduces the ABC-Hawkes algorithm for  parameter learning in the presence of distorted data. Section~\ref{sec_realdata} provides experimental results using Twitter data and simulations. We finish with a discussion of our work and contributions, as well as comments on future research. The \textsf{R} code for our analyses can be found in the supplementary material.

\section{Problem Overview} \label{sec_problemoverview}

In this section we define the Hawkes process and propose the implementation of distortion functions. We highlight the problem arising from missing or noisy events, which causes the Hawkes likelihood function to become computationally intractable.

\subsection{Hawkes Processes} \label{sec_Hawkes}

The (unmarked) Hawkes process introduced by \citet{hawkes_spectra_1971} is a self-exciting point process which models a collection of event times $Y = (t_i)_{i = 1}^N$. The occurrence of each event causes a short-term increase in the underlying point process conditional intensity function $\lambda(\cdot)$, known as self-excitation. This naturally produces temporal clusters of events, and it is hence an appropriate model for events which occur in bursts. More formally, a Hawkes process is a point process defined on the interval $[0, T]$  with a conditional intensity function: 
\begin{equation}
    \lambda(t| H_t,\theta) = \lambda_0(t|\theta) + \sum_{i: t > t_i}\nu(t-t_i|\theta)
    \label{eqn:hawkes}
\end{equation}
where $\theta$ is a vector of model parameters, and $H_t = \{t_i | t_i <t)$ denotes the set of events which occurred prior to time $t$. Here, $\lambda_0(t) > 0$ is the background intensity which defines the equilibrium rate at which events occur. The excitation kernel  $\nu(z)>0$ controls how much the intensity increases in response to an event at time $t$.  Typically, the excitation kernel is monotonically decreasing, so that more recent events are more influential. The choice of kernel varies from application to application. A typical choice for an unmarked point process is the exponential kernel \citep{NIPS2012_4834, rasmussen_bayesian_2013, Shelton}:
\begin{equation}
    \nu(z|\theta) = K \beta  e^{- \beta z}, \quad \theta = \{K, \beta\} \label{eq_exponential}.
\end{equation}

The Hawkes process can also be interpreted as a branching process \citep{hawkes_cluster_1974}, where ``immigrant'' events are  generated from the background process with intensity $\lambda_0(t)$, with each event spawning an additional off-spring process with intensity $\nu(z)$ which produces further events, and so on. 

For a set of observations $Y$ the likelihood of a Hawkes process is given by \citep{daley_introduction_2003}:
\begin{equation}
    p(Y | \theta)=\prod_{i=1}^N \lambda\left(t_{i} | \theta \right) e^{-\int_{0}^{\infty} \lambda\left(z | \theta \right) \,d z} \label{eq_etas_lik}
\end{equation}
where $\theta$ is the vector of unknown model parameters which must be estimated in order to fit the Hawkes process to the data. For ease of exposition, we will assume without loss of generality that the background intensity $\lambda_0(t) = \mu$ is constant, and that the excitation kernel is exponential. In this case,  $\theta =(\mu, K, \beta)$. However nothing in our algorithm requires these assumptions, and our method is equally applicable to any other choice of background rate or kernel function.

\subsection{Distorted Data} \label{sec_missing_data}

In many applications, the observed data will be distorted, i.e. noisy or containing  missing events. This is often due to data collection issues which result in some events being undetected or observed with error. Our proposal is to model this distorted data using a distortion function $h(\cdot)$, and now present some examples of specific choices of this function.

A common type of data distortion is \textbf{missing data} where a subset of the data is not observed. This is a well-known phenomenon which affects the use of Hawkes processes in earthquake forecasting since seismic detectors lose their ability to detect smaller earthquakes in the immediate aftermath of large earthquakes \citep{omi_estimating_2014}. In other examples, data from a certain period might simply have been lost, as is the case in the terrorism data set discussed by \cite{tucker_handling_2019}.

In the case of missing data, let $h(t)$ be a detection function specifying the probability that an event occurring at time $t$ will be successfully detected, and hence which will be present in the observed data $Y$. The observed data can be viewed as having arisen from the following generative process: First, a set of events $(t_1, \ldots, t_K)$ are generated from a Hawkes process with some intensity function $\lambda(\cdot)$. Then, for each event $t_k$ for $k = 1, \dots, K$, let $D_k = 1$ with probability $h(t_k)$ and $D_k = 0$ with probability $1-h(t_k)$. The observed data is then the collection of events for which $D_k = 1$, so that $Y = \{t_k | k: D_k = 1\}$. Note that this formulation of missing data is quite general and includes the classic ``gaps in the data'' scenario as a special case, since this is equivalent to setting $h(t) = 0$ during the period where gaps occur, and $h(t)=1$ elsewhere. However, this specification is highly flexible and also covers scenarios where events are not guaranteed to be missing for a certain period of time, but are instead missing with a (possibly time-dependent) non-zero probability, as in the earthquake scenario.

When data is potentially missing, it is very challenging to learn the $\theta$ parameter of the Hawkes process. If the Hawkes process did not have a self-exciting component, then the likelihood function in the presence of missing data would be obtained by assuming that the data came from a modified point process with intensity $r(t) = \lambda(t)h(t)$, i.e. the product of the intensity function and the detection function. The parameters of $r(\cdot)$ could then be learned using a standard method such as maximum likelihood or Bayesian inference. However, when working with Hawkes processes, the situation is substantially more complex. The main issue is that undetected events will still contribute to exciting the process intensity, i.e. the summation in Equation~\eqref{eqn:hawkes} needs to be over both the observed and unobserved events. The resulting likelihood function hence depends on both the set of observed events $Y$ and the set of unobserved events which we denote by $Y_{u}$. This requires integrating out the unobserved events, to give the likelihood function:
\begin{align}
    p(Y  | \theta) \propto     
    \int  p(Y, Y_{u} | \theta) \,
 \prod_{t_i \in Y} h(t_i) \prod_{t_l \in Y_{u}} (1- h(t_l)) \, d Y_{u} \label{eqn:missing}
\end{align}
where  $Y_u = \{t_k | k: D_k = 0\}$ denotes the set of missing events. Due to the integral over the unknown number of missing events, this likelihood function is intractable and cannot be evaluated. 

For \textbf{noisy data} (rather than missing data), we use a similar approach except that the distortion function $h(t)$ will now specify  the time at which an event is observed to have occurred, give that it truly  occurs at time $t$. For example, $h(t) = t + \varepsilon_t$ where $\varepsilon_t \sim N(0,\sigma^2_t)$ would be appropriate in a setting where the observation times are corrupted by Gaussian noise, while  $h(t) = t + c$ for a constant scalar $c$ would be used when there is a fixed delay present in the recording of all observation times. Such a synchronization example is studied by \citet{trouleau_learning_2019}. As in the case of missing data, this leads to an intractable likelihood function since the excitation in the summation from Equation~\eqref{eqn:hawkes} depends on the true observation time $t_i$ which is unknown. 

The distortion function may be parameterized by a vector of parameters $\xi$ which can denote, for example, the detection probability in the case of missing data, or the noise variance $\sigma_t$ in the case of noise. In some situations these parameters will be known due to knowledge of the underlying machinery used to detect the events. In other situations, they may be learned from the data.

Since all the above data distortion scenarios make direct learning of the Hawkes process impossible due to the intractable likelihood, we now propose a novel learning scheme for Hawkes processes with distorted data based on Approximate Bayesian Computation which we refer to as ABC-Hawkes.

\section{Approximate Bayesian Computation} \label{sec_ABC}

ABC is a widely studied approach to Bayesian inference in models with intractable likelihood functions \citep{beaumont_approximate_2002, marjoram_markov_2003,marin_approximate_2012}. We will now review the general ABC framework and then present a version of ABC for sampling from the posterior distribution of the Hawkes process parameters in the presence of distorted data.

Bayesian inference for a parameter vector $\theta \in \Theta$ assumes the existence of a prior $\pi(\theta)$ and a likelihood function $p(Y | \theta)$ for data $Y \in \mathcal{Y}$, with parameter inference based on the resulting posterior distribution $\pi(\theta | Y)$. Traditional tools for posterior inference such as Markov Chain Monte Carlo (MCMC) require the evaluation of the likelihood function and therefore cannot be applied when the likelihood is intractable. In this case, however, it may still be possible to generate samples $Y^{(j)}$ from the model. ABC is an approach to parameter and posterior density estimation which only uses such generated samples, without any need to evaluate the likelihood \citep{beaumont_approximate_2002}. %It is hence a suitable technique to perform inference for the missing data Hawkes process with an intractable likelihood. 

The core idea of ABC is as follows: We assume that the observed data $Y$ has been generated by some (unknown) value of $\theta$. For a proposed value of $\theta^{(j)}$, we generate pseudo-data $Y^{(j)}$ from the model $p(Y|\theta^{(j)})$ in a way which does not involve evaluating the likelihood function. If $\theta^{(j)}$ is close to the real $\theta$, then we would also expect $Y^{(j)}$ to be `close' to the real (observed) data $Y$, as measured by similarity function. We can hence accept/reject parameter proposals $\theta^{(j)}$ based only on the similarity between $Y$ and $Y^{(j)}$ .

This algorithm crucially depends on how we measure similarity between data sets $Y$ and $Y^{(j)}$. Typically, low-dimensional summary statistics $S(\cdot)$ of the data are chosen and then compared based on some distance metric $\mathcal{D}(\cdot ,\cdot )$ \citep{fearnhead_constructing_2012}. A proposed parameter $\theta^{(j)}$ is then accepted if this distance is less than a chosen threshold $\epsilon$. Generally, this procedure will not target the true posterior $\pi(\theta | Y)$, but instead targets the ABC posterior $\pi_{ABC}(\theta | \mathcal{D}(S(Y) , S(Y^{(j)}) < \epsilon)$. However, if the statistics $S(\cdot)$ are chosen to be the sufficient statistics for the model parameters and $\epsilon \to 0$, then $\pi_{ABC}(\theta | \mathcal{D}(S(Y) , S(Y^{(j)}) < \epsilon) \to \pi(\theta | Y)$ \citep{marin_approximate_2012}. A direct implementation of this ABC procedure is based upon rejection sampling, where $\theta^{(j)}$ is proposed from the prior distribution \citep{pritchard_population_1999}. However this can be inefficient, so instead ABC-MCMC methods can be used which make proposals based on a Metropolis-Hastings kernel without the need to evaluate the likelihood function, which can lead to a higher acceptance rate for ABC \citep{beaumont_approximate_2002}.

In most realistic applications, it is not possible to choose a low-dimensional set of summary statistics which is sufficient for the parameter vector, since only the limited class of distributions that lie in the exponential family admit a finite dimensional set of sufficient statistics \citep{brown_fundamentals_1986}. Therefore, finding suitable summary statistics and an appropriate distance metrics for a particular model is non-trivial, and is a vital part of designing efficient ABC algorithms \citep{marin_approximate_2012}. Some authors construct summary statistics which are carefully tailored to their application \citep{aryal_fitting_2019}, while others present a semi-automatic way to constructing summary statistics \citep{fearnhead_constructing_2012}. \citet{bernton_approximate_2019} develop a general approach, that uses the Wasserstein Distance between the observed and synthetic data set, hence eliminating the need for summary statistics altogether.

\subsection{ABC-Hawkes Algorithm} \label{sec_ABC_pointprocess}

Our core idea is to use ABC to perform inference for the Hawkes process with distorted data, which circumvents the intractability of the likelihood function. This is made possible by the fact that simulating data from the distorted generative model is straightforward, and can be done by first simulating data from the Hawkes process with intensity function $\lambda(\cdot)$ which represents the true (unobserved) data and then distorting these events based on the distortion function $h(\cdot)$ as discussed in Section~\ref{sec_missing_data}, which gives the observed data $Y$. The simulation from $\lambda(\cdot)$ can be carried out using a standard simulation algorithm for Hawkes process such as the thinning procedure of \citet{ogata_lewis_1981}. The distortion of the data is then performed by applying $h(\cdot$) to each simulated data point, which introduces missingness and/or noise. For example in the case of gaps, this would consist of deleting the simulated observations which lie within the gap region, while for noise it would consist of adding on noise to the simulated data, as specified by $h(\cdot)$. The resulting observations are hence a realization of the Hawkes process with intensity function $\lambda(\cdot)$ that has been distorted through $h(\cdot)$.

For ABC-Hawkes we use a variant of the ABC-MCMC algorithm, as shown in Algorithm~\ref{alg_abc_MCMC}. This is an extension of the usual Metropolis-Hastings algorithm which essentially replaces the intractable likelihood function with an estimate based on the simulated data and can be shown to converge correctly to the ABC posterior  $\pi_{ABC}(\theta | \mathcal{D}(S(Y) , S(Y^{(j)}) < \epsilon)$ \citep{marjoram_markov_2003}.

\begin{algorithm}[tbp]
   \caption{ABC-Hawkes} \label{alg_abc_MCMC}
 \begin{algorithmic}
   \STATE {\bfseries Input:} observed data $Y$ where data is distorted according to a distortion function $h(\cdot)$,  the parameter prior $\pi(\cdot)$, the desired number of posterior samples $J$, a function to compute the $P$ summary statistics $S_1(\cdot),\ldots,S_P(\cdot)$, the $P$ threshold levels $\epsilon_p > 0$, and a Metropolis-Hastings transition kernel $q(\cdot | \cdot)$ \\
   \hrulefill
   \STATE{Initialise $\theta^{(0)}$}
   \FOR {$j=1$ {\bfseries to} $J$} 
        \STATE{$\theta^* \sim q(\cdot | \theta^{(j -1)})$}
        \STATE{$ Z^{*} \sim p(\cdot | \theta^{*})$, where $p(\cdot)$ simulates from a Hawkes process}
        \STATE{$Y^{*} = h(Z^{*})$}
        \IF{$|S_p(Y^{*}) - S(Y)| < \epsilon_p$ for all $p = 1, \dots, P$}
            \STATE{With probability $min \left\{1, \frac{q(\theta^{(j-1)}|\theta^{*}) \pi(\theta^*)}{q(\theta^*|\theta^{(j-1)}) \pi(\theta^{(j-1)})}\right\}$ set $\theta^{(j)} = \theta^*$}
        \ELSE
        \STATE{Set $\theta^{(j)} = \theta^{(j-1)}$}
        \ENDIF
   \ENDFOR 
   \STATE {\bfseries Output:} $(\theta^{(1)}, \dots, \theta^{(J)})$
  \end{algorithmic}
\end{algorithm}

The choices of summary statistics $S(Y)$ and corresponding threshold $\epsilon$ are crucial for the application of ABC \citep{marin_approximate_2012}. While the ABC literature offers a wealth of theory on summary statistics, their actual construction is less prominent and tends to be highly application-dependent. In the context of Hawkes processes, this is complicated by the non-i.i.d. structure of the data, which renders many existing approaches impossible as they require multiple (bootstrapped) samples of the original data set; such as using random forests to find summary statistics  \citep{pudlo_reliable_2016}, or utilizing a classifier or reinforcement learning to judge the similarity between data sets \citep{gutmann_likelihood-free_2018, li_learning_2018}. These approaches are not applicable here as only one data set is available and bootstrap sampling cannot be used due to the complex dependency structure of self-exciting point processes. 

 While there has been some previous literature on ABC for Hawkes processes outside of the distorted data setting \citep{ertekin_reactive_2015, shirota_approximate_2017} we found that the presented summary statistics did not extend well to distorted data. Instead, through extensive simulations we have found a set of summary statistics which we have empirically found to accurately captures the posterior distribution of the Hawkes process parameters, and in the Experiments section we will provide evidence to support this. For ABC-Hawkes, the resulting summary statistics  $S(Y) = (S_1(Y), \ldots , S_7(Y)))$ we use are:

\begin{itemize}
    \item $S_1(Y)$: The logarithm of the number of observed events in the process. This is  highly informative for the $\mu$ and $K$ parameters, since $\mu$ controls the number of background events, while $K$ defines how much total triggering is associated with each event.
    \item $S_2(Y)$: The median of the event time differences $\Delta_i = t_i - t_{i-1}$ divided by the mean event time differences $\mathbb{E}[\Delta]$. This is highly informative for the parameters of the self-excitation kernel (e.g. $\beta$, for an exponential kernel). 
    \item $S_3(Y) \dots S_5(Y)$: Ripley's $K$ statistic \citep{ripley_1977_modelling} for a window size of $(1, 2, 4)$. This counts the events that happen within the respective window-length of each other. We do not use a correction for the borders. This captures the degree of clustering in the event sequence and is hence informative for $K$ and the parameters of the excitation kernel.
    \item $S_6(Y), S_7(Y)$: The average of the $\Delta_i$ differences that lie above their $90\%$-quantile $\mathbb{E}[\Delta_i | \Delta_i > q_{90}]$ and below their median $\mathbb{E}[\Delta_i | \Delta_i < q_{50}]$. In extensive simulation studies we found that these complement the other statistics well and lead to an accurate approximation of the posterior distribution.
\end{itemize}

When using multiple summary statistics, careful consideration must be made when combining them together. We choose to work with a separate thresholds $\epsilon_p$ for each of the summary statistics. We set this to be a fraction of the empirical standard deviation of the summary statistic based on a pilot run of the simulation, such that $0.01-0.1$\% of the proposals are accepted \citep{vihola_use_2020}. We hence accept a proposed value $\theta^{(j)}$ only if $\mathcal{D}(S_p(Y), S_p(Y^{(j)}) < \epsilon_p$ for all $p$, using the absolute value function ($L_1$ norm) as the distance metric $\mathcal{D}(\cdot, \cdot)$.

In the case where the detection function $h(\cdot)$ is parameterized based on an unknown set of hyperparameters $\xi$, these can also be sampled from their posterior distribution at each stage in the above MCMC algorithm using standard Metropolis-Hastings methods. 

\section{Experimental Results} \label{sec_realdata}

We now present experimental results to show the performance of the ABC-Hawkes algorithm. First, we will present evidence that the set of summary statistics we presented above capture most of the information in the parameter posterior distribution for the Hawkes process when distorting is not present. Next, we will investigate how accurately they allow the parameters to be estimated in the presence of distortion. To do this, we will manually insert data distortion into a simulated event sequences where we have access to the true event times. This allows us to compare the posterior distribution estimated by ABC-Hawkes on the distorted data to the ``true'' posterior which would have been obtained if we had access to the undistorted data.
\subsection{No-distortion Setting}

We first confirm whether the above ABC summary statistics accurately allows the posterior distribution to be estimated in a standard Hawkes process without distortion. We will then compare our approach to that of \citet{ertekin_reactive_2015} which is the only existing example we could find of applying ABC to Hawkes processes, although they do not consider the data distortion setting.

\begin{figure}[t]
  \flushleft
   \begin{subfigure}{.09\textwidth}
      \flushright
      Data Set 1
  \end{subfigure}
  \begin{subfigure}{.28\textwidth}
      \centering
      \includegraphics[width=.99\linewidth]{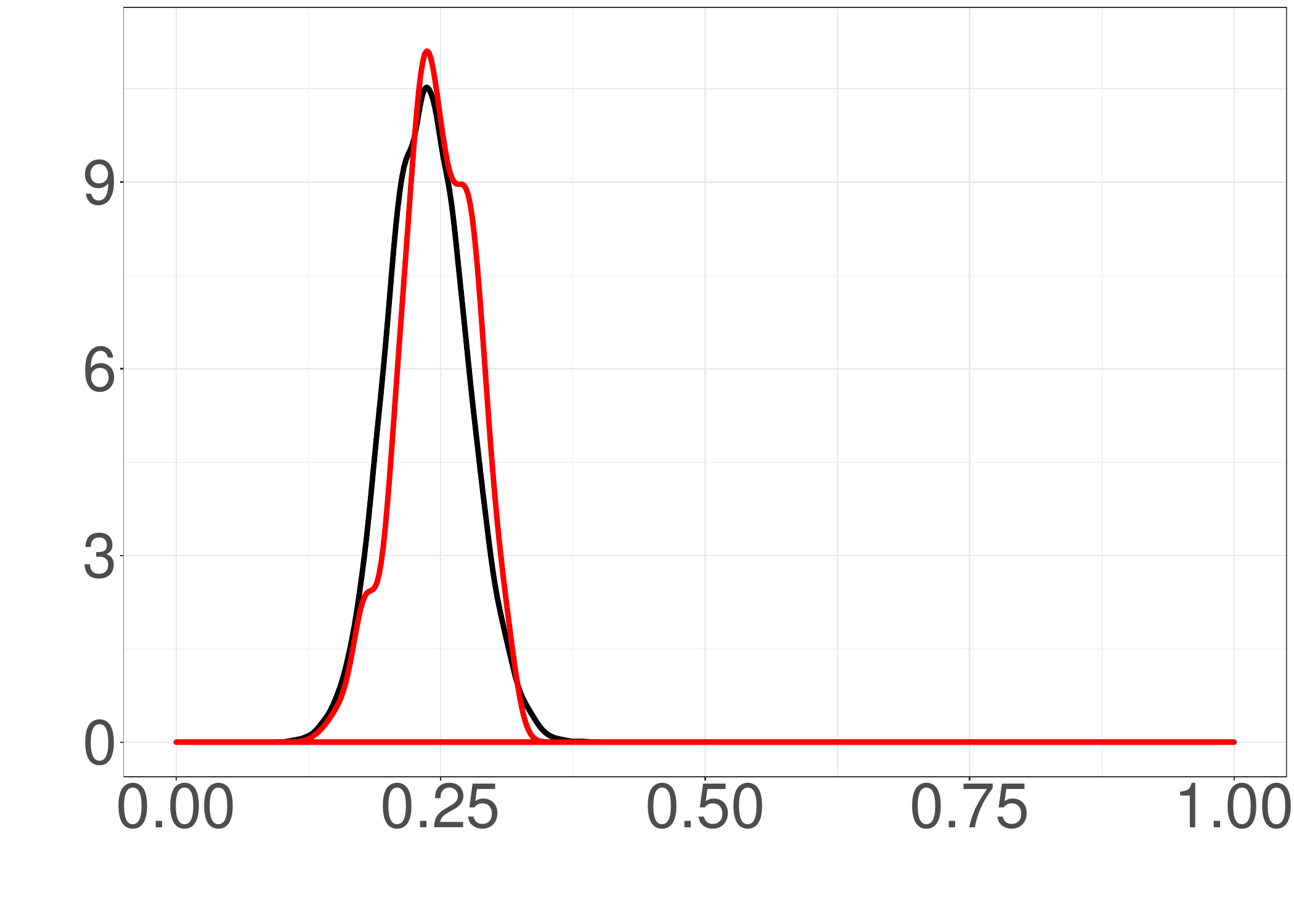}
  \end{subfigure}
  \begin{subfigure}{.28\textwidth}
      \centering
      \includegraphics[width=.99\linewidth]{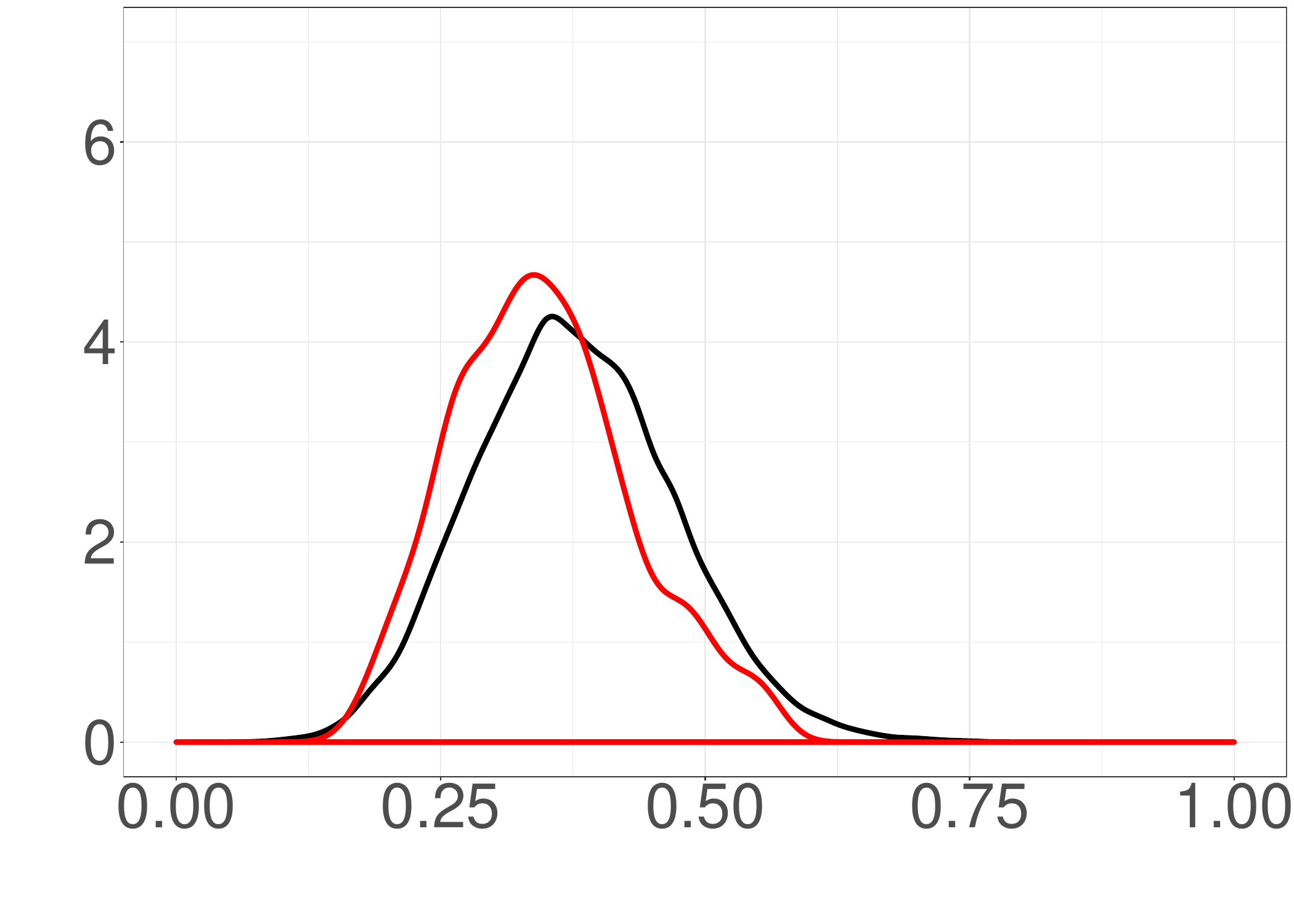}
  \end{subfigure}
  \begin{subfigure}{.28\textwidth}
      \centering
      \includegraphics[width=.99\linewidth]{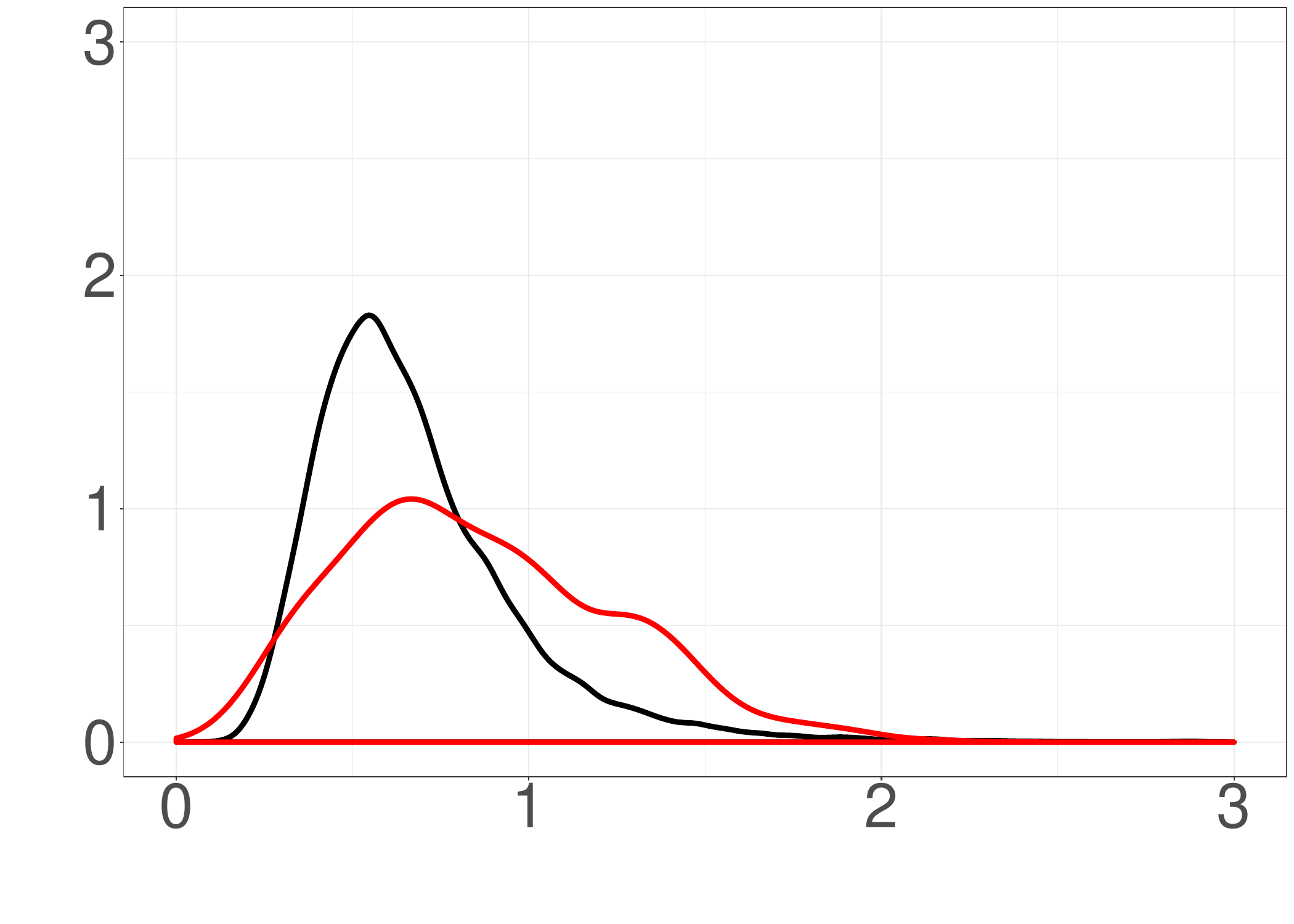}
  \end{subfigure} \\
  \begin{subfigure}{.09\textwidth}
      \flushright
      Data Set 2
  \end{subfigure}
  \begin{subfigure}{.28\textwidth}
      \centering
      \includegraphics[width=.99\linewidth]{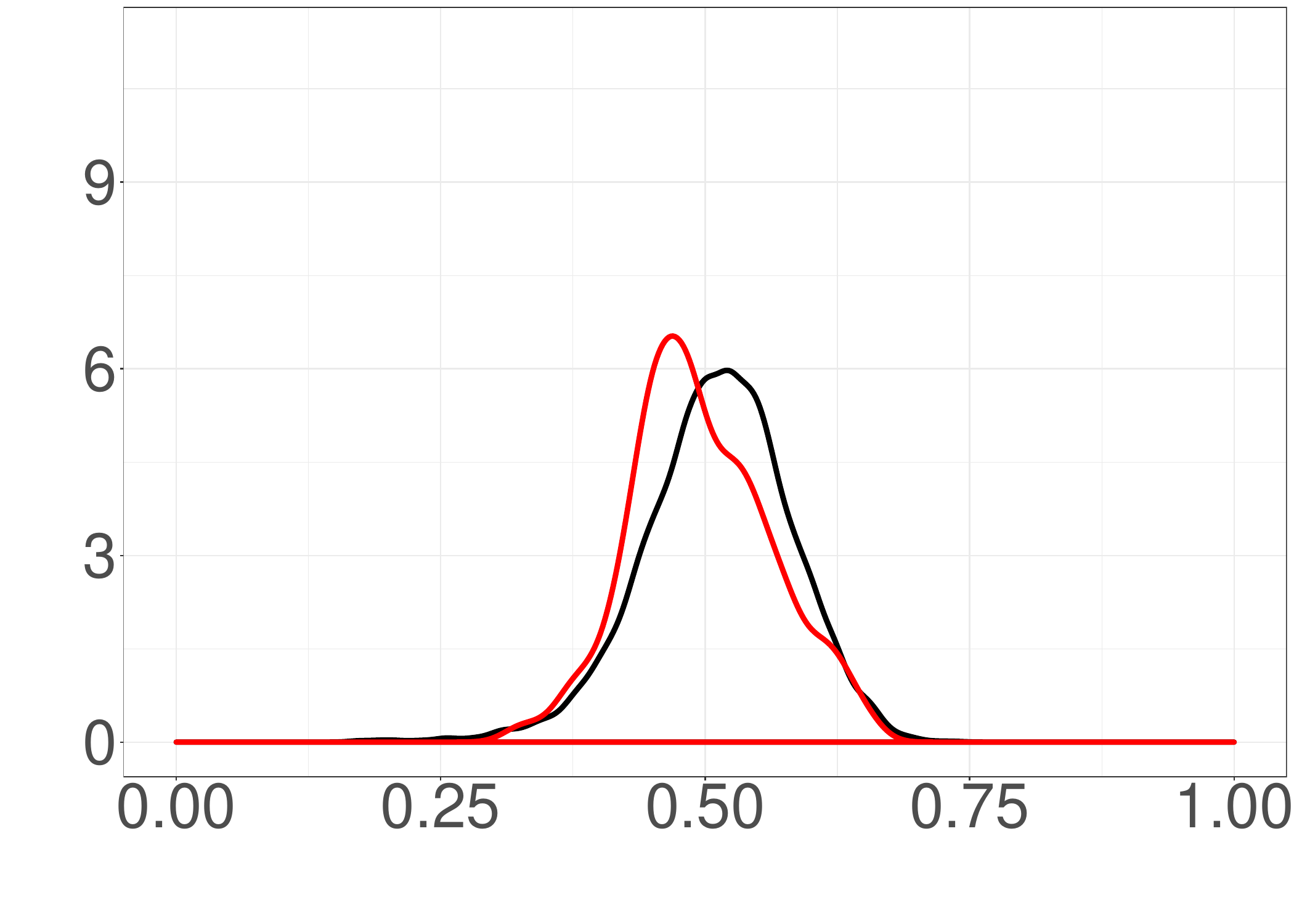}
  \end{subfigure}
  \begin{subfigure}{.28\textwidth}
      \centering
      \includegraphics[width=.99\linewidth]{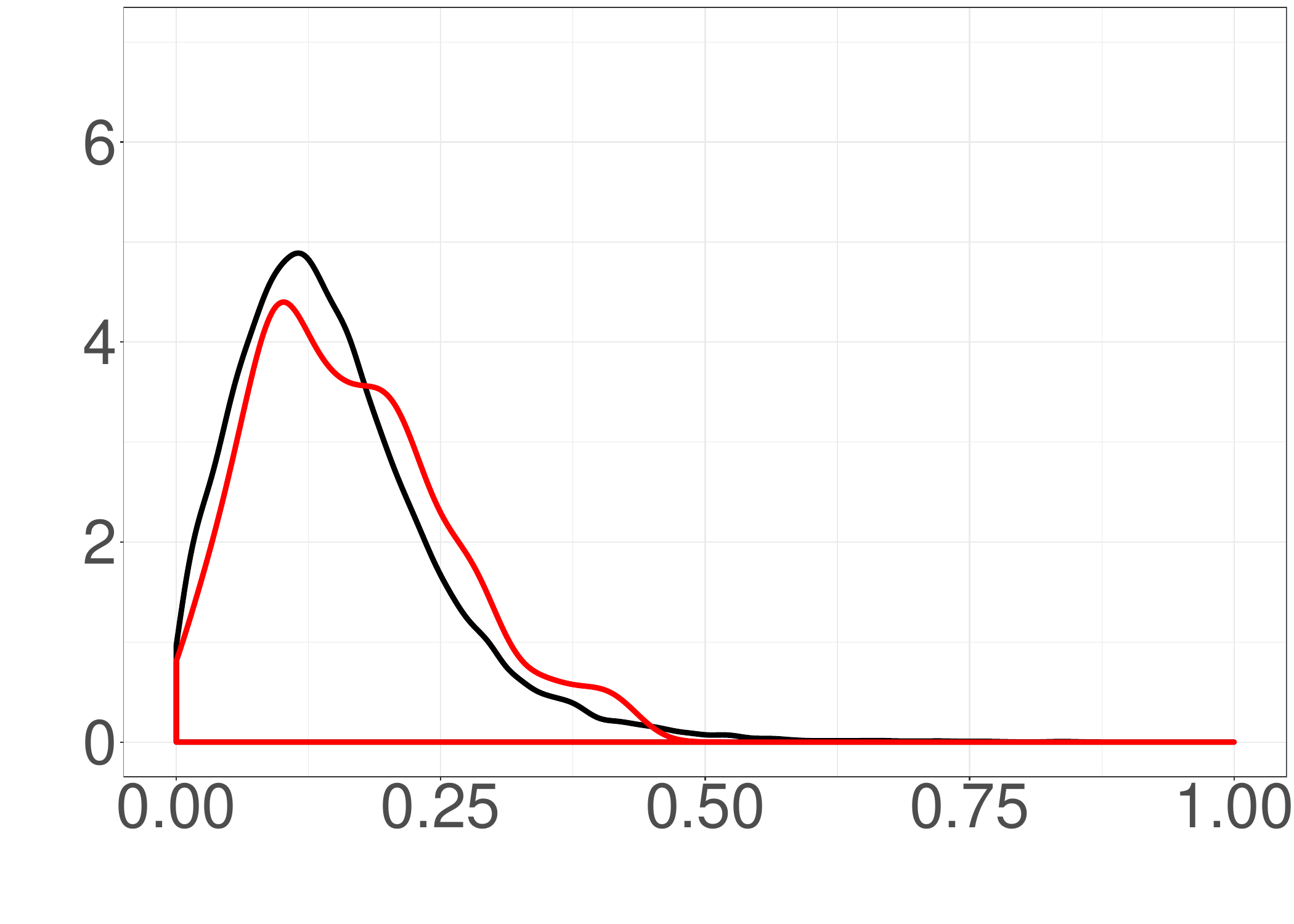}
  \end{subfigure}
  \begin{subfigure}{.28\textwidth}
      \centering
      \includegraphics[width=.99\linewidth]{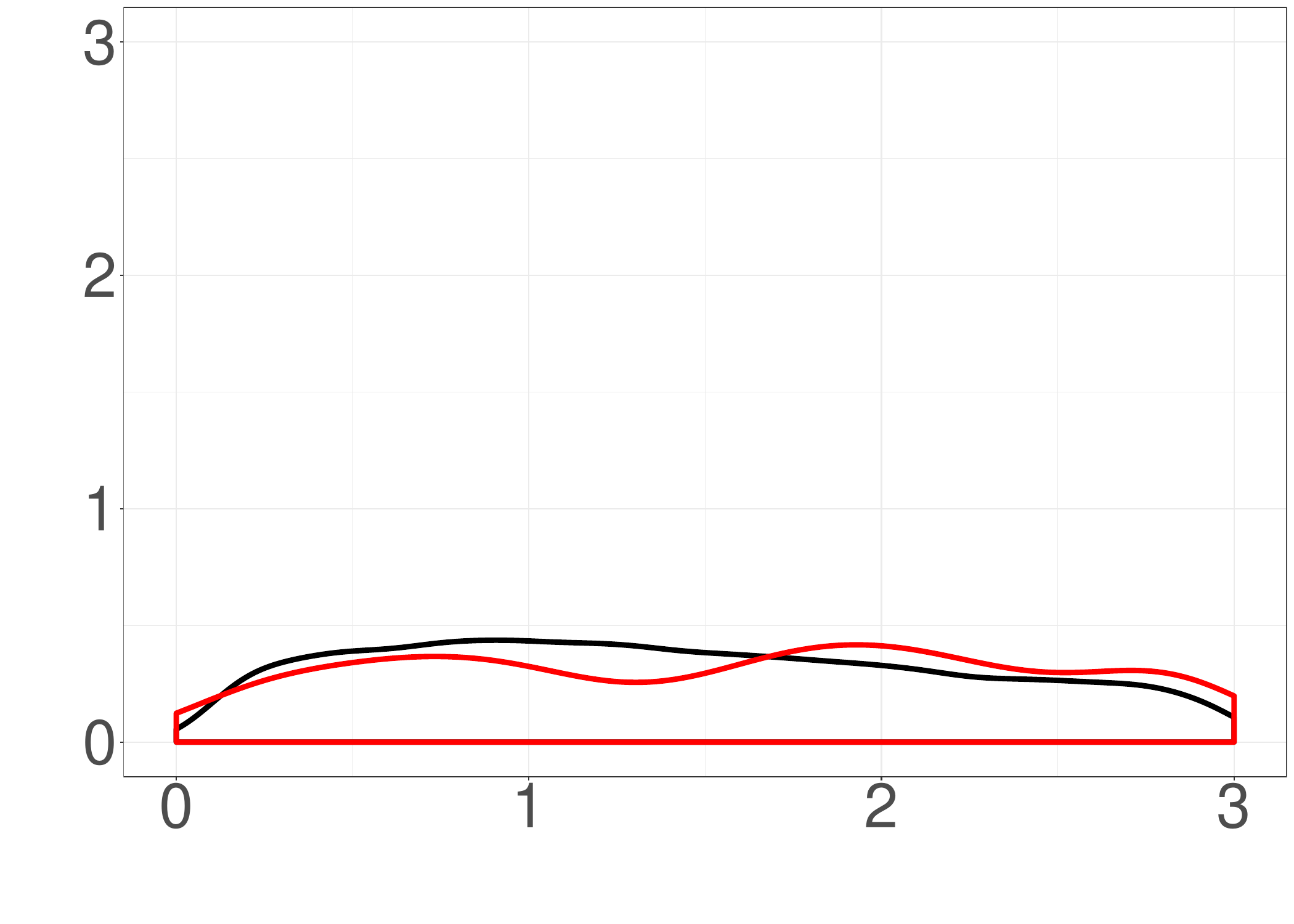}
  \end{subfigure}\\
  \begin{subfigure}{.09\textwidth}
      \flushright
      Data Set 3
  \end{subfigure}
  \begin{subfigure}{.28\textwidth}
      \centering
      \includegraphics[width=.99\linewidth]{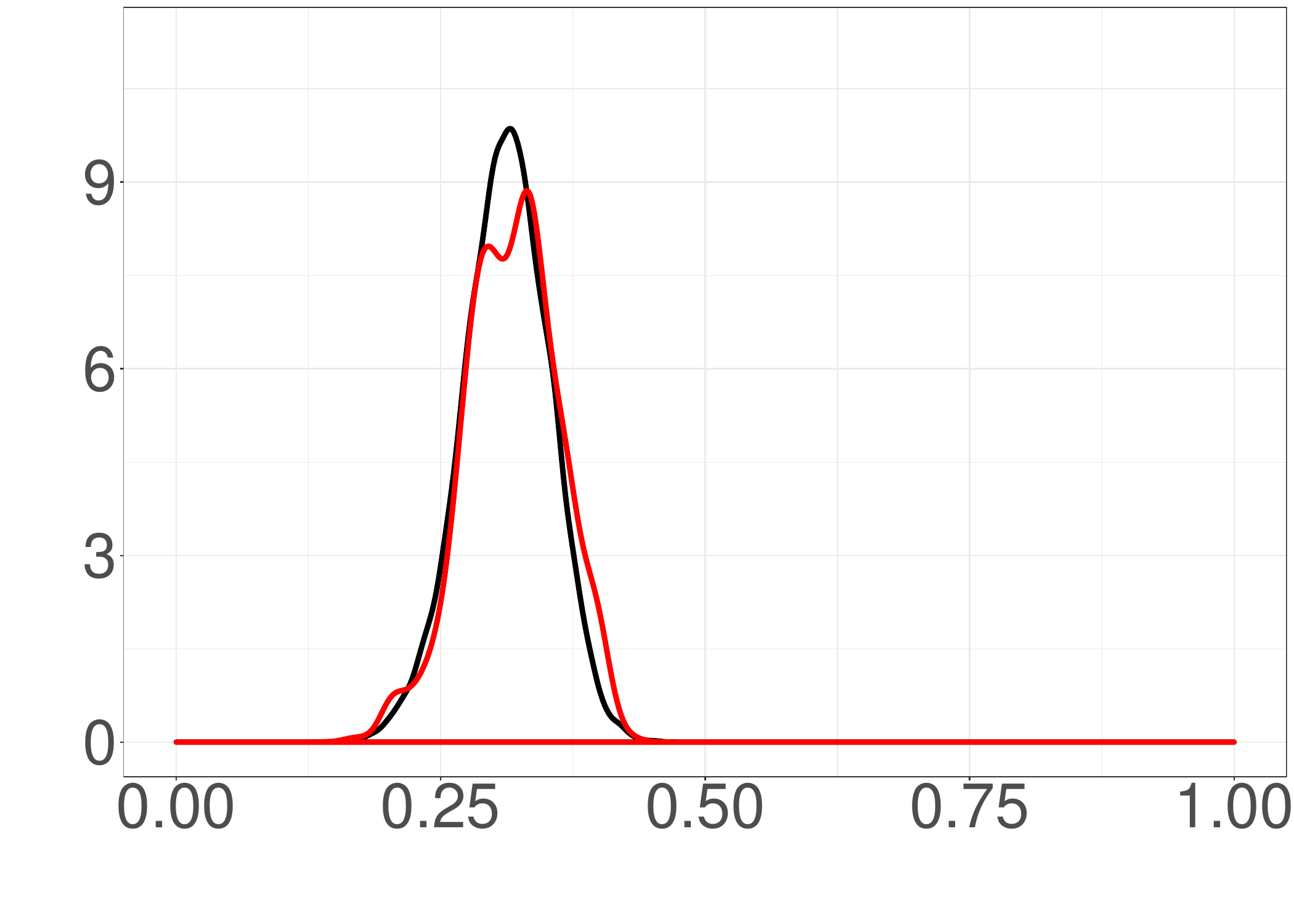}
      \caption{$\mu$}
  \end{subfigure}
  \begin{subfigure}{.28\textwidth}
      \centering
      \includegraphics[width=.99\linewidth]{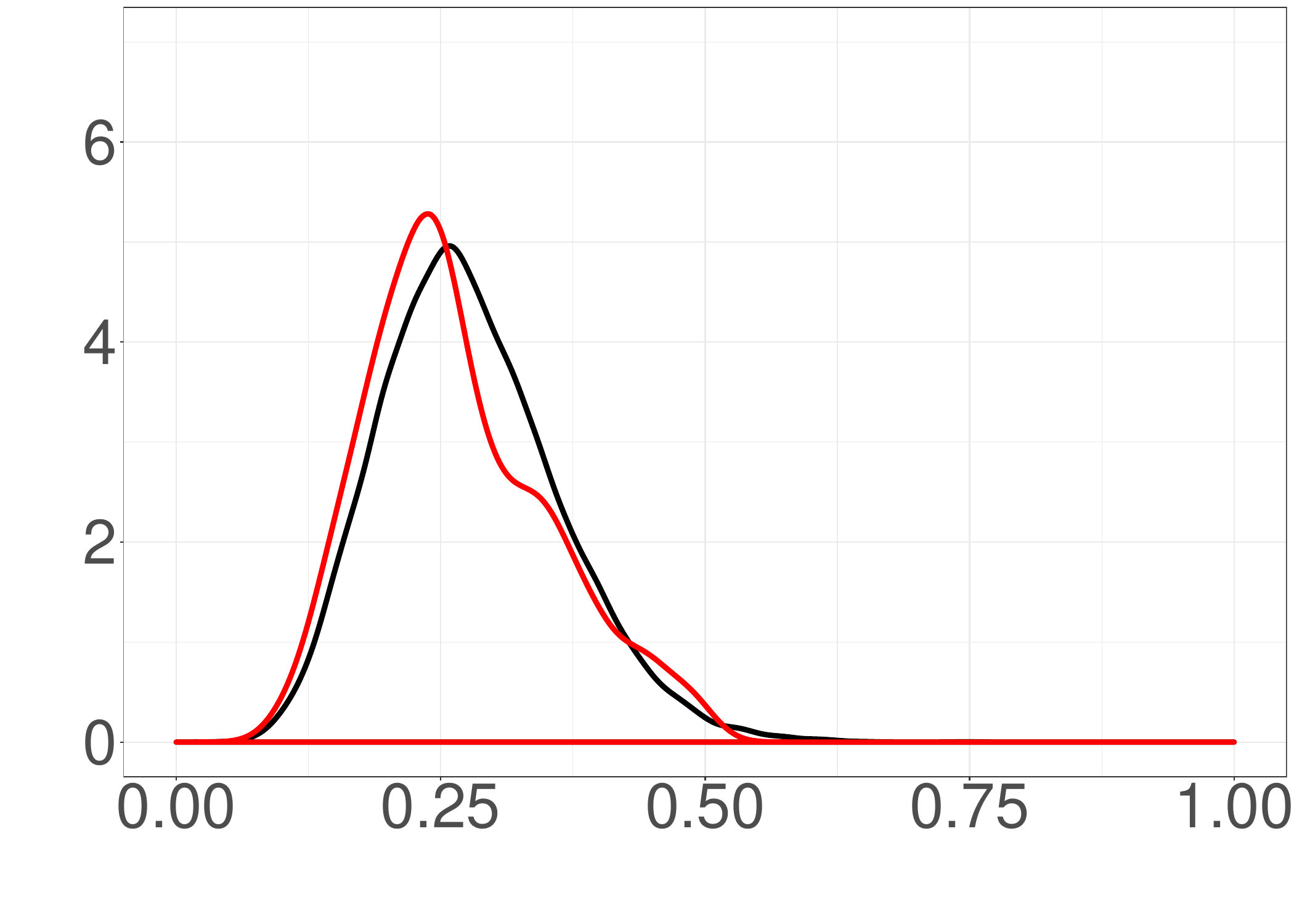}
      \caption{$K$}
  \end{subfigure}
  \begin{subfigure}{.28\textwidth}
      \centering
      \includegraphics[width=.99\linewidth]{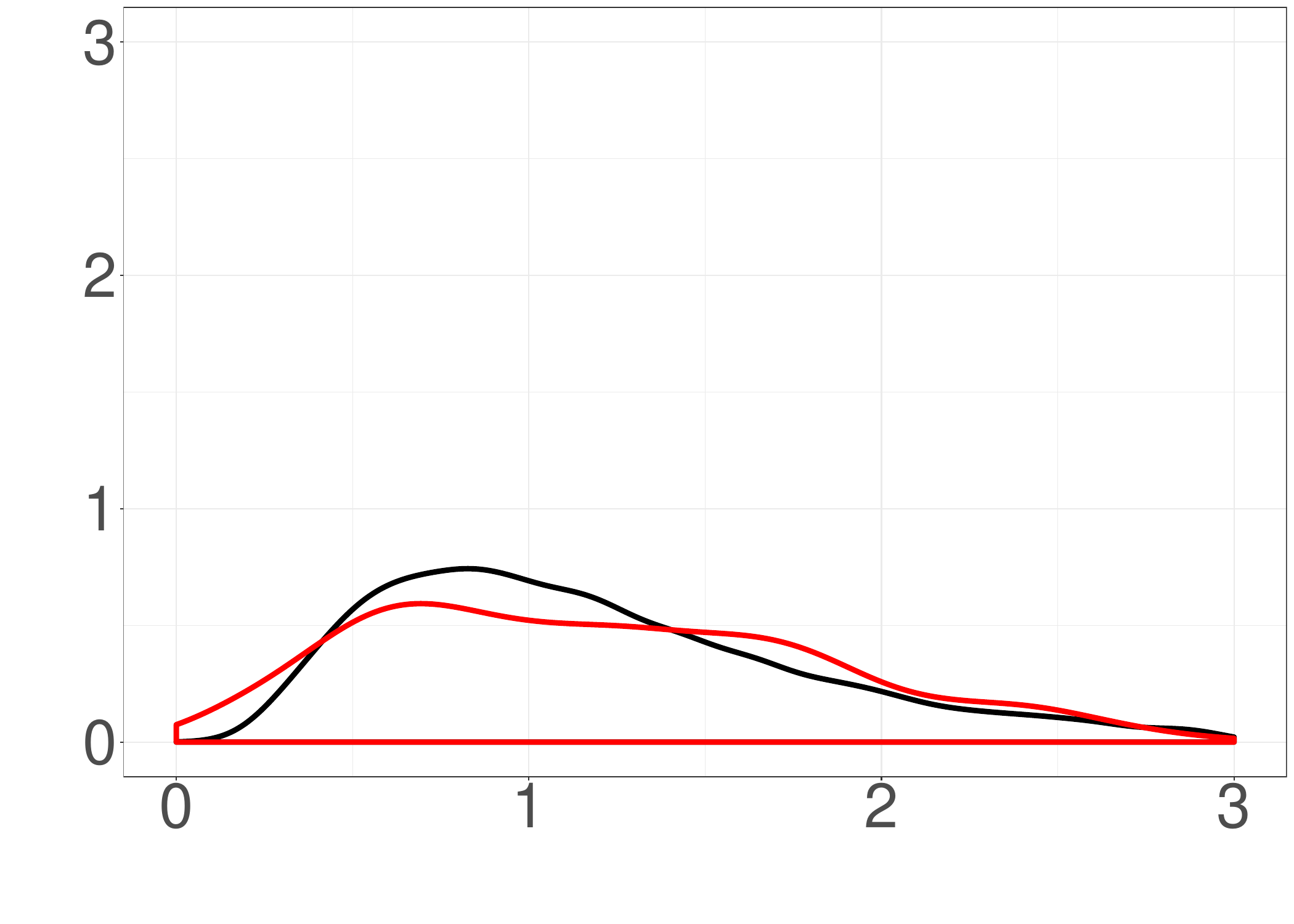}
       \caption{$\beta$}
  \end{subfigure}
  \caption{Undisturbed data posterior distributions. Each row uses a different simulated data set. Parameters $(\mu, K, \beta)$ are chosen as $(0.2, 0.5, 0.5)$, $(0.4, 0.3, 0.8)$, and $(0.3, 0.3, 1)$ for the three data sets. The black curve shows the true posterior estimated by samples generated from Stan, red represents ABC-Hawkes. All estimates are based on the complete, undistorted data sets.}
  \label{fig_post_proofofconcept}
\end{figure}

To investigate the capabilities of our algorithm to recover the true posterior distribution without distortion we generate three data sets and compare the posterior estimates from ABC-Hawkes to the ground truth from Stan. We use the following priors: $\mu \sim \mathcal{U}(0.05, 0.85)$, $ K \sim \mathcal{U}(0, 0.9)$, $ \beta \sim \mathcal{U}(0.1, 3)$. As shown in Figure~\ref{fig_post_proofofconcept} ABC-Hawkes can approximate the posterior distributions, both in location and shape, in this simulation study. We note that the overestimation of the posterior variance for $\beta$ in the first data set is an often observed issue in ABC stemming from a necessary non-zero choice of the $\epsilon_p$ threshold \citep{li_convergence_2017}.

In \citet{ertekin_reactive_2015}, a Hawkes process is defined which has an intensity function containing both an inhibiting and exciting component, as well as a constant background intensity $\lambda_0$, and $C_1$, a term to deal with zero-inflation. In their simulation study (provided in their supplementary materials) they fix both $\lambda_0$ and $C_1$ to their true values, which only leaves two free parameters, i.e. they do not consider the full estimation problem where the background intensity needs to be learned in addition to the triggering kernel. To estimate the posterior distributions using ABC \citet{ertekin_reactive_2015} use two summary statistics: the log-number of events and the KL divergence between the histograms of the interevent times $\Delta_i$ of the true and simulated data set. 

\begin{figure}[t]
  \flushleft
   \begin{subfigure}{.2\textwidth}
      \flushright
      Data Set 1
  \end{subfigure}
  \begin{subfigure}{.3\textwidth}
      \centering
      \includegraphics[width=.99\linewidth]{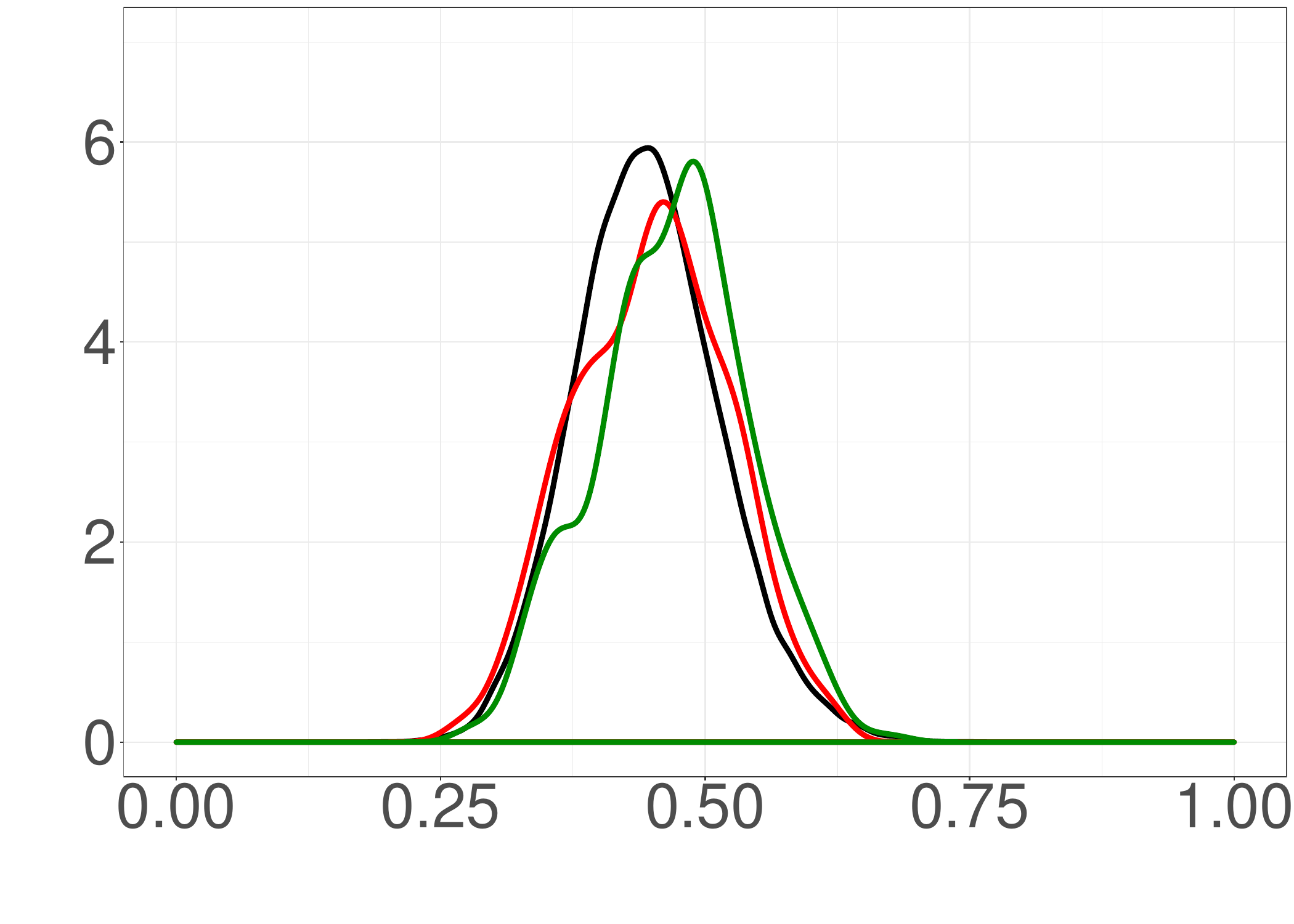}
  \end{subfigure}
  \begin{subfigure}{.3\textwidth}
      \centering
      \includegraphics[width=.99\linewidth]{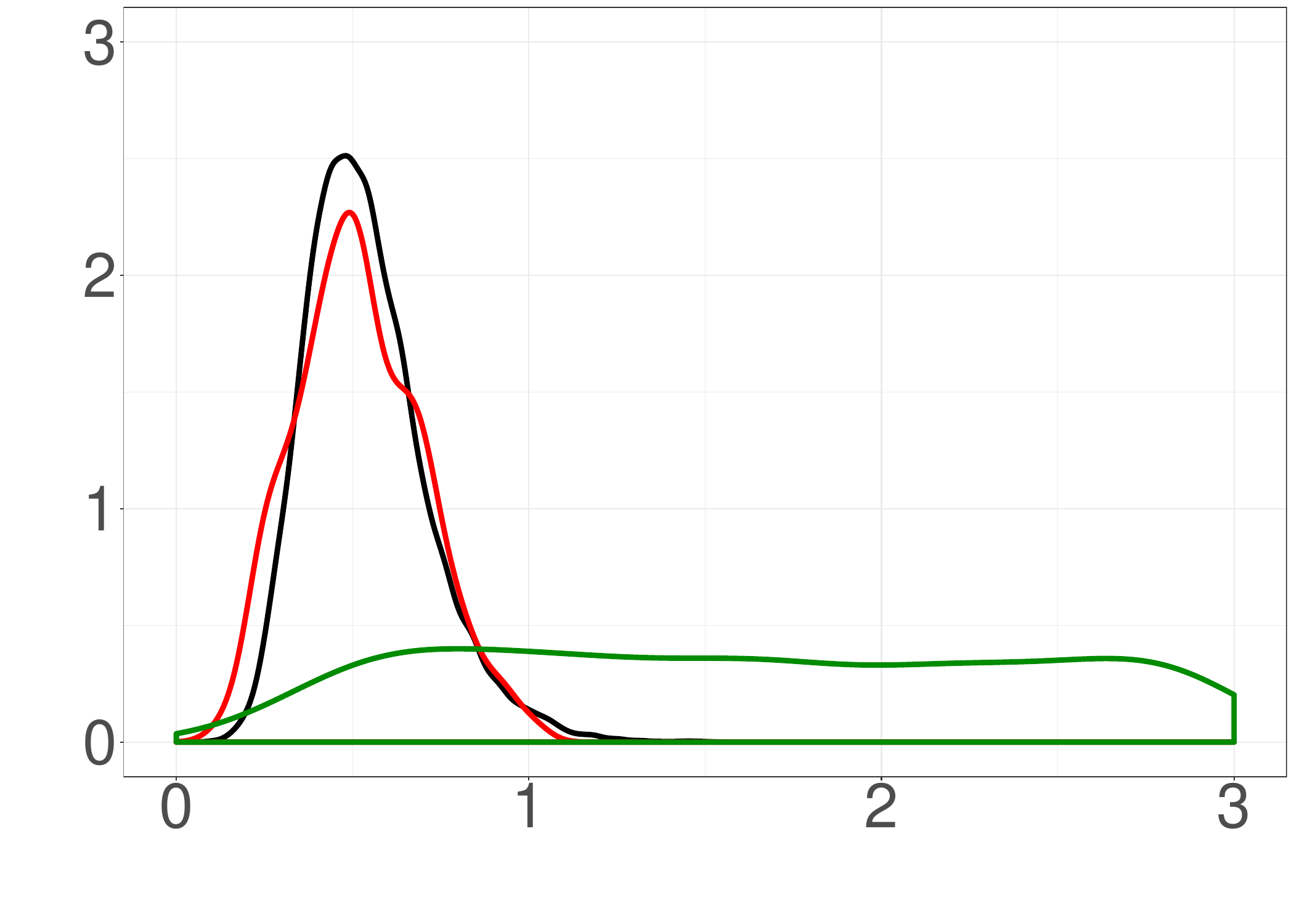}
  \end{subfigure} \\
  \begin{subfigure}{.2\textwidth}
      \flushright
      Data Set 2
  \end{subfigure}
  \begin{subfigure}{.3\textwidth}
      \centering
      \includegraphics[width=.99\linewidth]{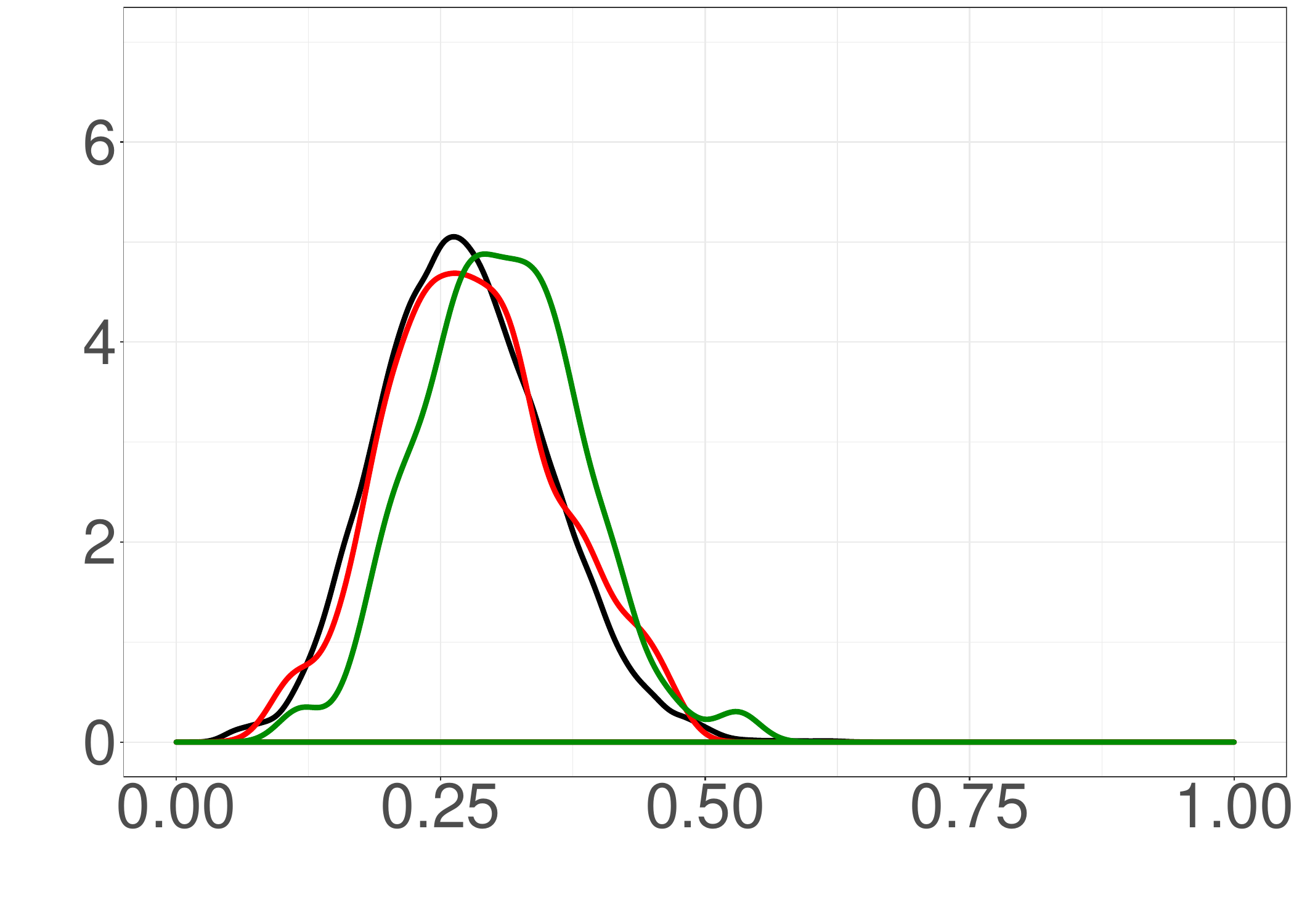}
  \end{subfigure}
  \begin{subfigure}{.3\textwidth}
      \centering
      \includegraphics[width=.99\linewidth]{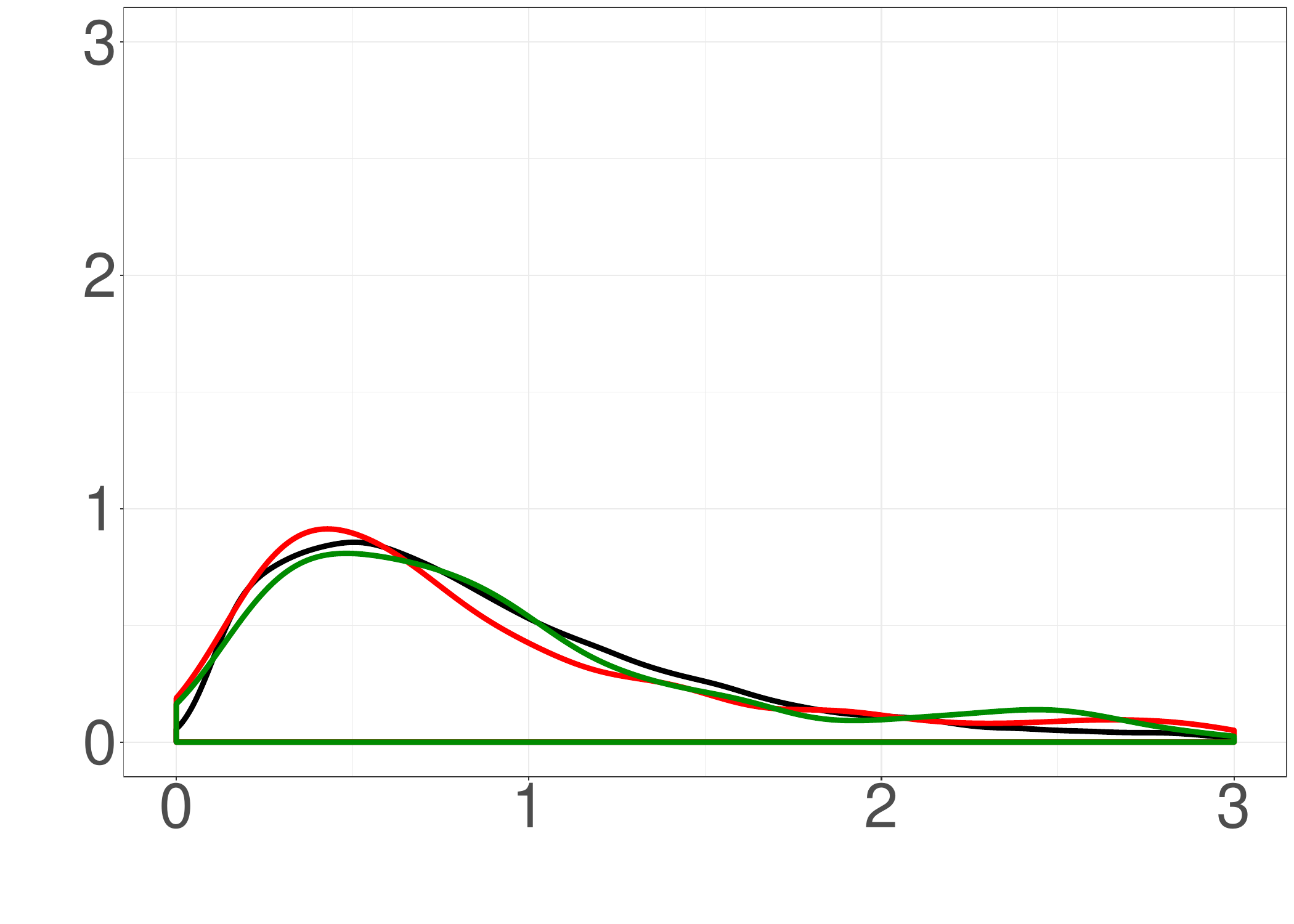}
  \end{subfigure}\\
  \begin{subfigure}{.2\textwidth}
      \flushright
      Data Set 3
  \end{subfigure}
  \begin{subfigure}{.3\textwidth}
      \centering
      \includegraphics[width=.99\linewidth]{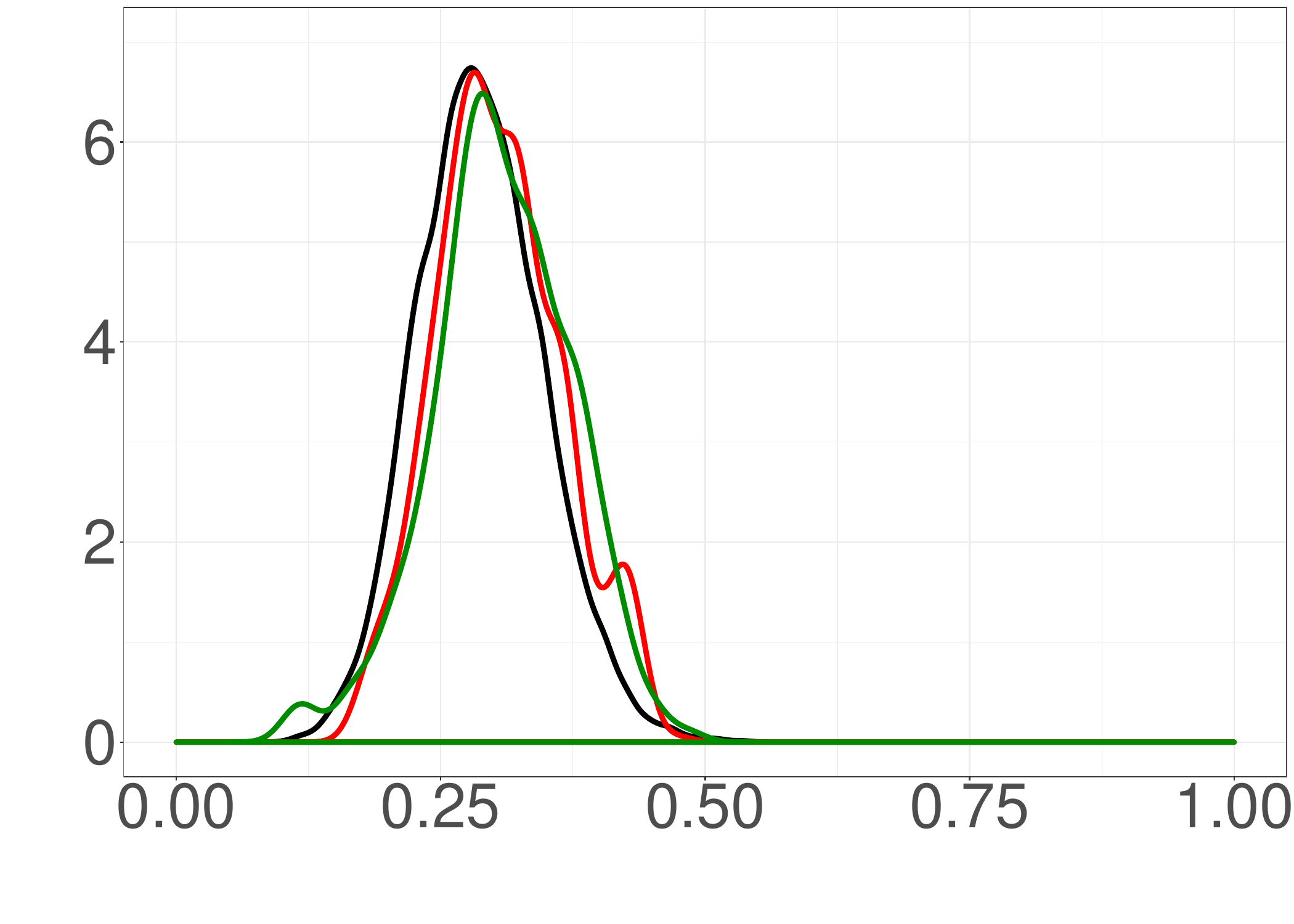}
      \caption{$K$}
  \end{subfigure}
  \begin{subfigure}{.3\textwidth}
      \centering
      \includegraphics[width=.99\linewidth]{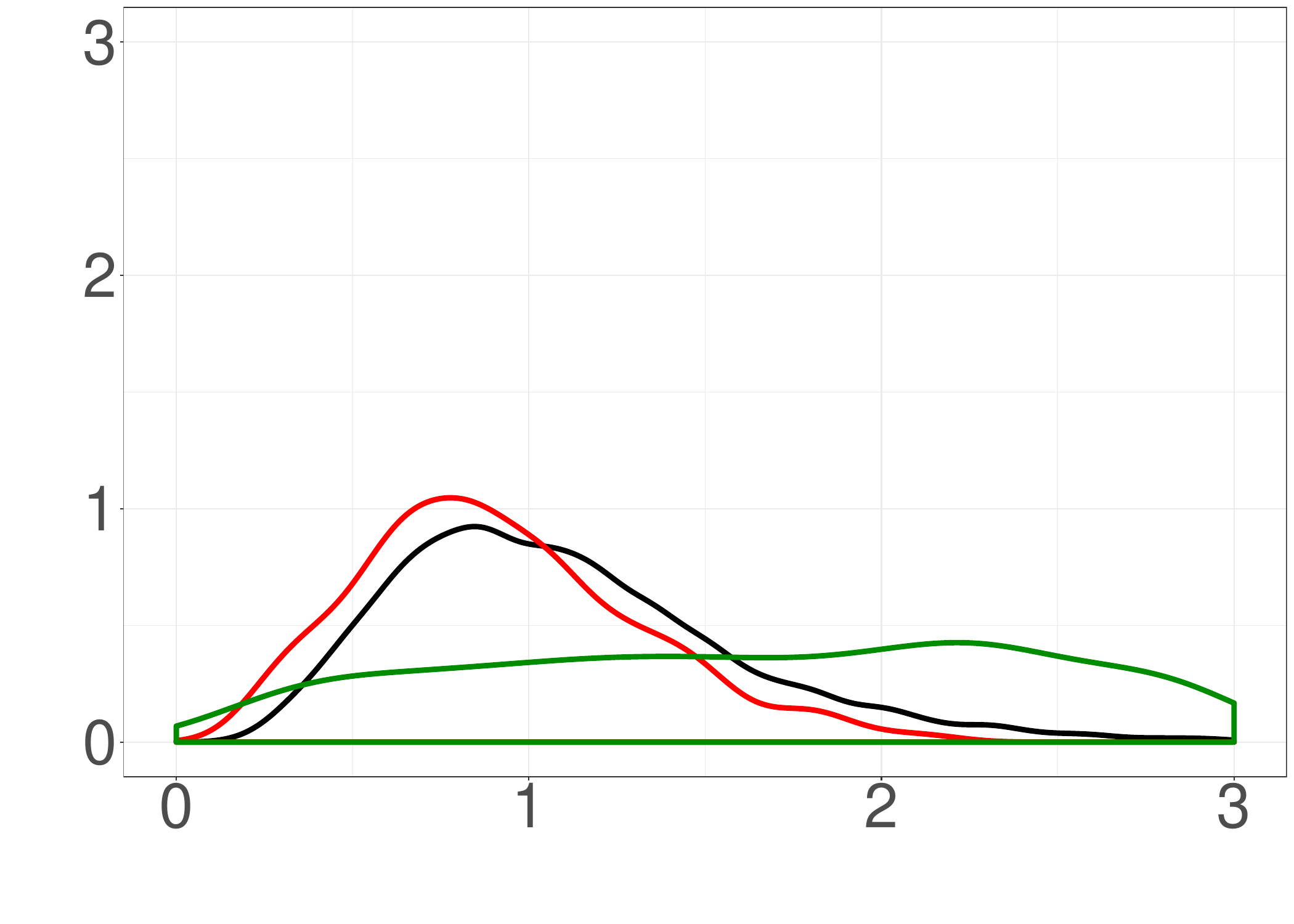}
       \caption{$\beta$}
  \end{subfigure}
  \caption{Undisturbed data posterior distributions. Each row uses a different simulated data set. Parameters $(K, \beta)$ are chosen as $(0.5, 0.5)$, $(0.3, 0.8)$, and $(0.3, 1)$ for the three data sets. The black curve shows the true posterior estimated by samples generated from Stan, red represents ABC-Hawkes, and green the approach by \citet{ertekin_reactive_2015}. All estimates are based on the complete, undistorted data sets.}
  \label{fig_post_ert}
 
\end{figure}

For our comparison we generate three data sets from a Hawkes process. We use a constant background intensity $\mu$, which is assumed fixed and known for all methods to facilitate a direct  comparison with \citet{ertekin_reactive_2015}. Hence, we are left to estimate the posteriors of the two free parameters from the exponential excitation kernel from Equation~\eqref{eq_exponential}, hence $\theta' = (K, \beta)$. The priors for the Hawkes parameters are chosen to be relatively uninformative: $ K \sim \mathcal{U}(0, 0.9)$, $ \beta \sim \mathcal{U}(0.1, 3)$. Without any data distortion, Figure~\ref{fig_post_ert} compares the true posterior distribution to the estimates using the two summary statistics from \citet{ertekin_reactive_2015} and ABC-Hawkes. It is evident that ABC-Hawkes does a better job at capturing the posterior distributions across all parameters and data sets, showing that it is able to  estimate the Hawkes parameters despite not evaluating the likelihood function.

\subsection{Distortion -- Twitter Data Example} \label{sub_twitter}

We next apply the ABC-Hawkes algoorithm to parameter estimation in distorted data. For this purpose, we will manually insert distortion into a real data set and then compare the posterior distribution estimated from the distorted data, to that estimated from the original undistorted data.

For the real data, we choose to study the occurrence time of tweets on Twitter, which previous research has shown can be accurately modeled by a Hawkes process \citep{mei_neural_2017}. We use a Twitter data set that was previously analyzed by \citet{rizoiu_tutorial_2017} and describes the retweet cascade of an article published in the New York Times. While this data set is complete (and we can hence obtain the true posterior distribution), we will manually create a gap in the data to assess whether the true posterior can be recovered using only this distorted data.

To evaluate ABC-Hawkes, we use the first $150$ event times in the tweet data. To artificially create a gap, we delete all observations from observation $t_{60}$ to observation $t_{90}$, to produce the incomplete data. The be relatively uninformative priors are chosen as above: $\mu \sim \mathcal{U}(0.05, 0.85)$, $ K \sim \mathcal{U}(0, 0.9)$, $ \beta \sim \mathcal{U}(0.1, 3)$. The restriction that $K<1$ is standard, and ensures that data sampled from the Hawkes process contain a finite number of events with probability 1. 

We generate samples from the true parameter posteriors using Markov Chain Monte Carlo as implemented in the Stan \citep{Stan_RStan_2019} probabilistic programming language, as applied to the complete data. This represents the idealized ``ground truth'' posterior that would be obtained if we had access to the true undistorted data. We then  apply ABC-Hawkes to the observed data only (i.e. the distorted, incomplete data), with the goal of recovering this true posterior. In the implementation of the ABC-MCMC algorithm we use independent random walk Gaussian proposal distribution for the $q(\cdot|\cdot)$ transition kernels, with  standard deviations $(0.05, 0.05, 0.2)$ for $(\mu, K, \beta)$ respectively. To assess the performance of ABC-Hawkes, we compare the obtained posterior to three alternative methods that could be used: (1) MCMC (using Stan) applied to the incomplete data, which represents the naive attempt to learn the Hawkes parameters directly using only the observed data and ignoring the missing data. (2) MCMC (using Stan) applied only to the observations $[0, T_a]$ before the start of the gap. This is ``unbiased'' since it uses a sequence of data where all tweets are available, but is inefficient due to the smaller resulting set. (3) The missing data algorithm suggested by \cite{tucker_handling_2019}, which is a specialised algorithm applicable only to data where the distortion consists of gaps. Their approach alternates between two steps. First, they impute the missing data based on the complete pre-gap data and a given parameter vector. Second, they update the parameter vector based on the imputed data set of full length, where likelihood evaluations are possible. This approach has the advantage of targetting the correct posterior, however it can suffer from slow mixing since the probability of a MCMC example being accepted is low.

We note that approaches (2) and (3) are only applicable to this specific choice of distortion function where the distortion consists of no detected events at all during the gap period, and (unlike ABC-Hawkes), are not applicable to more general types of distortion, as will be discussed in the next section.

\begin{figure}[t]
  \centering
  \begin{subfigure}{.3\textwidth}
      \centering
      \includegraphics[width=.99\linewidth]{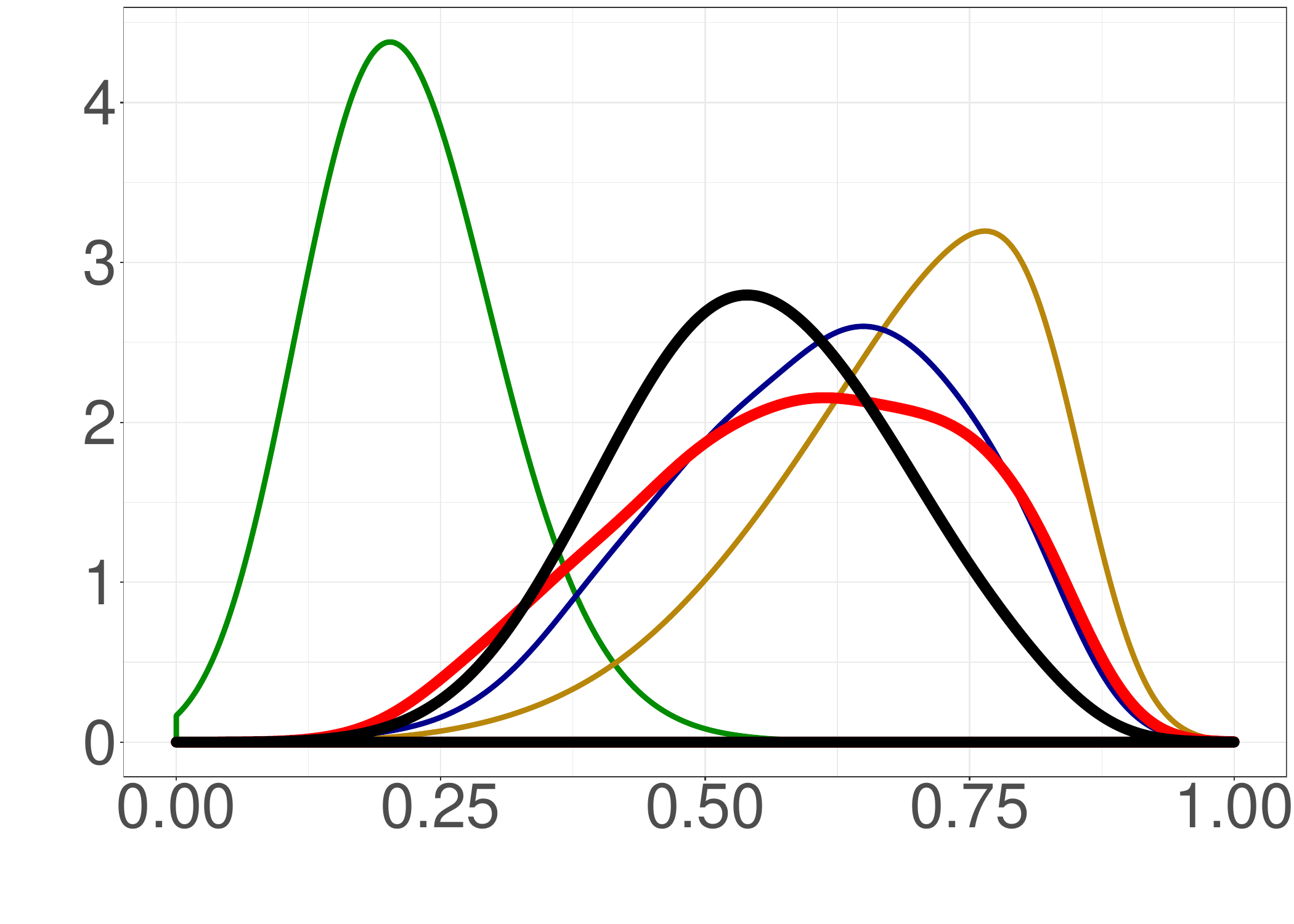}
      \caption{$\mu$}
      \label{subfig_unmarked_mu}
  \end{subfigure}
  \begin{subfigure}{.3\textwidth}
      \centering
      \includegraphics[width=.99\linewidth]{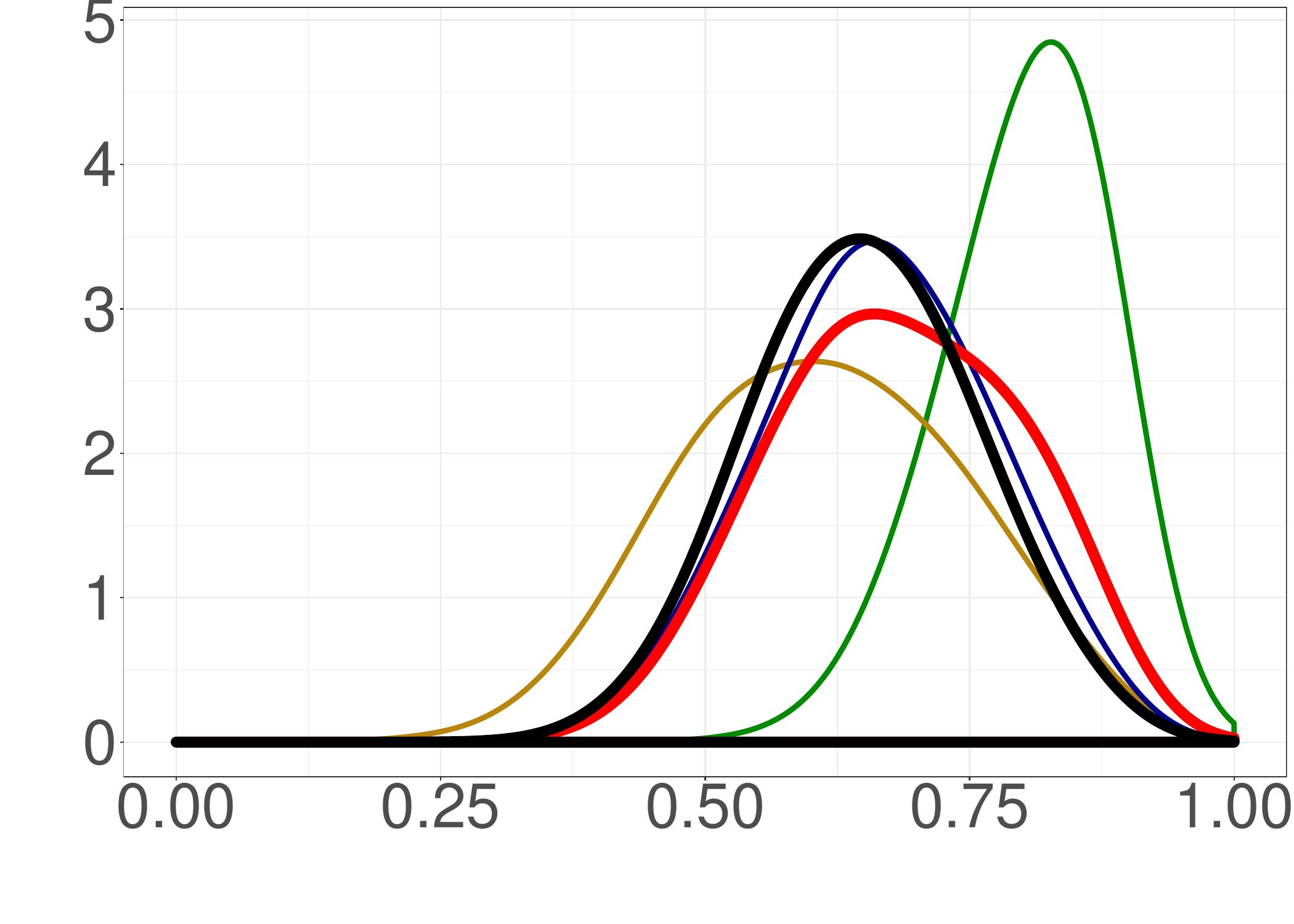}
      \caption{$K$}
      \label{subfig_unmarked_K}
  \end{subfigure}
  \begin{subfigure}{.3\textwidth}
      \centering
      \includegraphics[width=.99\linewidth]{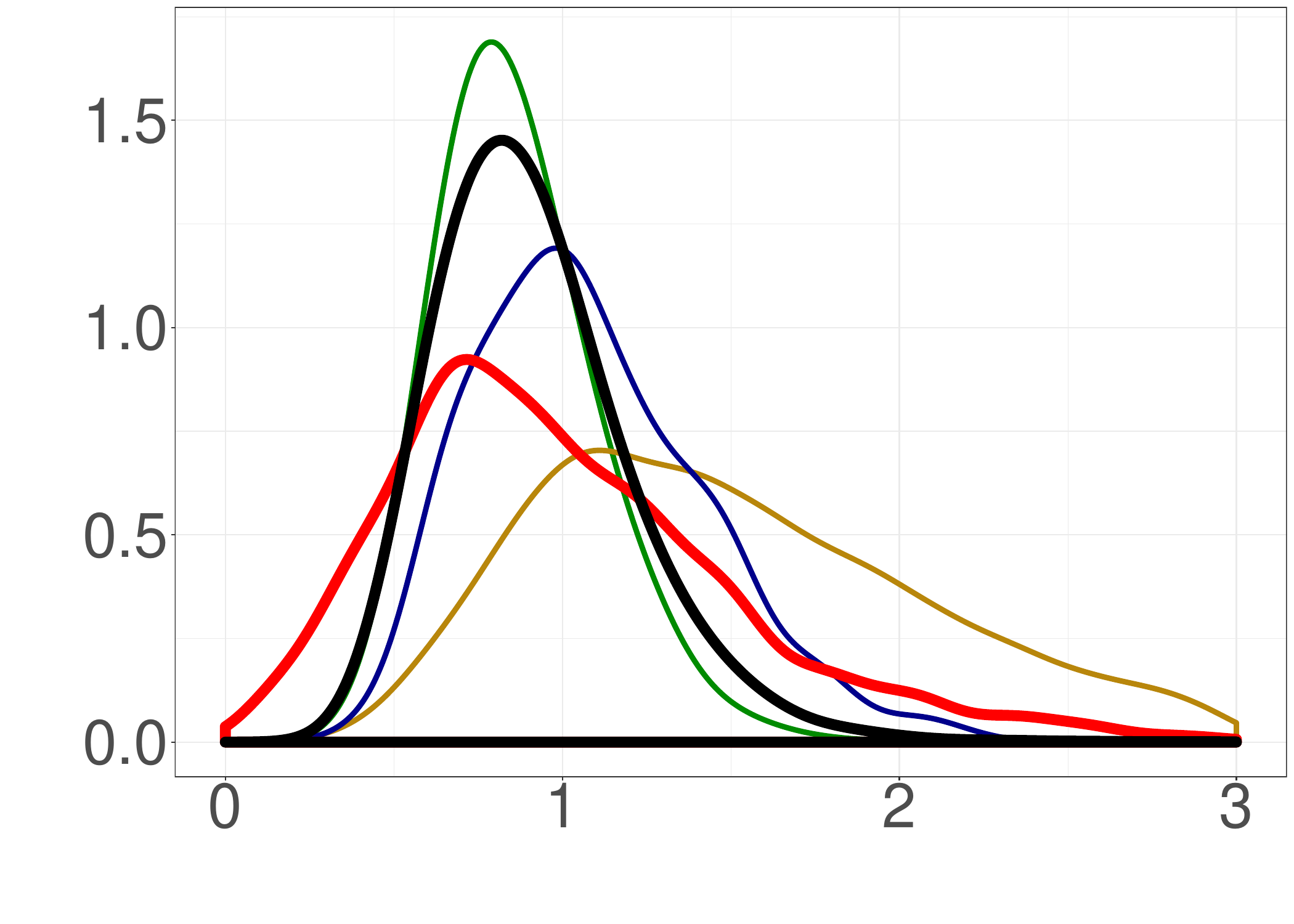}
      \caption{$\beta$}
      \label{subfig_unmarked_beta}
  \end{subfigure}
  \caption{Twitter data posterior distributions. Black represents the true posterior using both observed and missing data. Red represents ABC-Hawkes using only the incomplete data. The naive approach using only the incomplete data is green, while the yellow and blue lines respectively show the approach that only uses the observations before the gap, and the  \citet{tucker_handling_2019} imputation method.}
  \label{fig_post_unmarked}
  
\end{figure}

Figure~\ref{fig_post_unmarked} shows the resulting posterior density estimates for all these methods, and Table~\ref{tab_twitter} contains the posterior means and standard deviations. It can be seen that the ABC-Hawkes algorithm does an excellent job of recovering the posterior distribution despite the missing data, and is very close to the true posterior mean for each of the model parameters. In contrast, the naive approach (which ignores the gap) produces a highly biased posterior which is not close to the true posterior means of $\mu$ and $K$. Both the approach from \cite{tucker_handling_2019} and the method which uses only the observations prior to the gap do substantially better than the naive approach, but are inferior to ABC Hawkes. 

%We next compare the ability of ABC Hawkes to predict how many events will occur in the future, given the observed data. We define the test period to be the interval $[T,T_{test}]$ where $T$ is the upper bound of the training set. In actuality, there were 25 observations during this period.  For each of the above methods including ABC Hawkes, we simulated 1000 data sets on $[T,T_{test}]$ using the estimated parameters, with excitation coming from the observed events in the training set.

%The four methods are then used to predict the number of observations in the test period

%which contains $25$ observations. For each method, we simulate $1000$ data sets on $[T, T_{test}]$ using the incomplete training data set \citep[with imputed points for the model by][]{tucker_handling_2019} as previous observations. ABC-Hawkes estimates an average of $35.58$ observations (with standard deviation $\sigma = 20.43$), the complete on the incomplete data set $23.73$ ($\sigma = 22.02$) and on the complete subset $37.7$ ( $\sigma = 21.55$), and the method by \citet{tucker_handling_2019} $36.32$ ($\sigma = 20.29$). Here, the (biased) method using the incomplete data set profits from the fact that the retweets diminish over time, which indicates a non-constant $\mu$.  

\begin{table}[t]
  \caption{Twitter data posterior mean (and standard deviation)}
  \label{tab_twitter}
  \centering
  \begin{tabular}{lccc}
\toprule
Model & $\mu$ & $K$ & $\beta$ \\ 
  \midrule
  \textbf{True Posterior} &  \textbf{0.55 (0.13)} &\textbf{0.65 (0.10)} & \textbf{0.91 (0.28)} \\
  \midrule
 ABC-Hawkes  &  0.59 (0.15) & 0.68 (0.11) & 1.00 (0.51) \\
 Naive & 0.22 (0.08) & 0.80 (0.07) & 0.87 (0.23) \\ 
Pre-gap only  & 0.68 (0.13) & 0.61 (0.13) & 1.49 (0.57) \\ 
\citet{tucker_handling_2019}  & 0.61 (0.14) & 0.66 (0.10) & 1.08 (0.34) \\ 
 \addlinespace[0.5ex] 
\bottomrule
\end{tabular}
\end{table}

\subsection{Other Distortion Settings}

The above section showed that ABC-Hawkes can accurately recover the true posterior when there are gaps in the data. However a key advantage of our approach is that \citep[unlike the approach from][]{tucker_handling_2019} it can also be used when the data is distorted in other ways. To investigate this, we use a simulation study where we generate data sets from a Hawkes process and manually distort it in various ways. The first type of distortion involves a time-varying detection rate, while the second involves noisy data. The prior distribution for $\theta$ is taken to be the same as above and we again used Stan \citep{Stan_RStan_2019} to sample from the idealized parameter posterior using the undistorted data, which acts as a baseline that would be obtained if no distortion were present.

For the first type of distortion, we use a linearly decaying detection function where an event that occurs at time $t$ is observed with probability $  h(t) = 1- (a + b \, \frac{t}{T})$ where $a = 0.35$ and $b = -0.25$, and missing otherwise. Hence, earlier events have a lower probability of being observed. Similar time-varying detection functions have been applied in the earthquake literature \citep{ogata_immediate_2006} and hence this is plausible specification. For the second type of distortion, we create a ``noisy'' version of the data set by adding Gaussian errors to each observation, i.e. replacing each $t_i$ with $t'_i = t_i + \varepsilon_i$ where $\varepsilon_i \sim \mathcal{N}(0,0.5^2)$.

\begin{figure}[!th]
  \centering
  \begin{subfigure}{.3\textwidth}
      \centering
      \includegraphics[width=.99\linewidth]{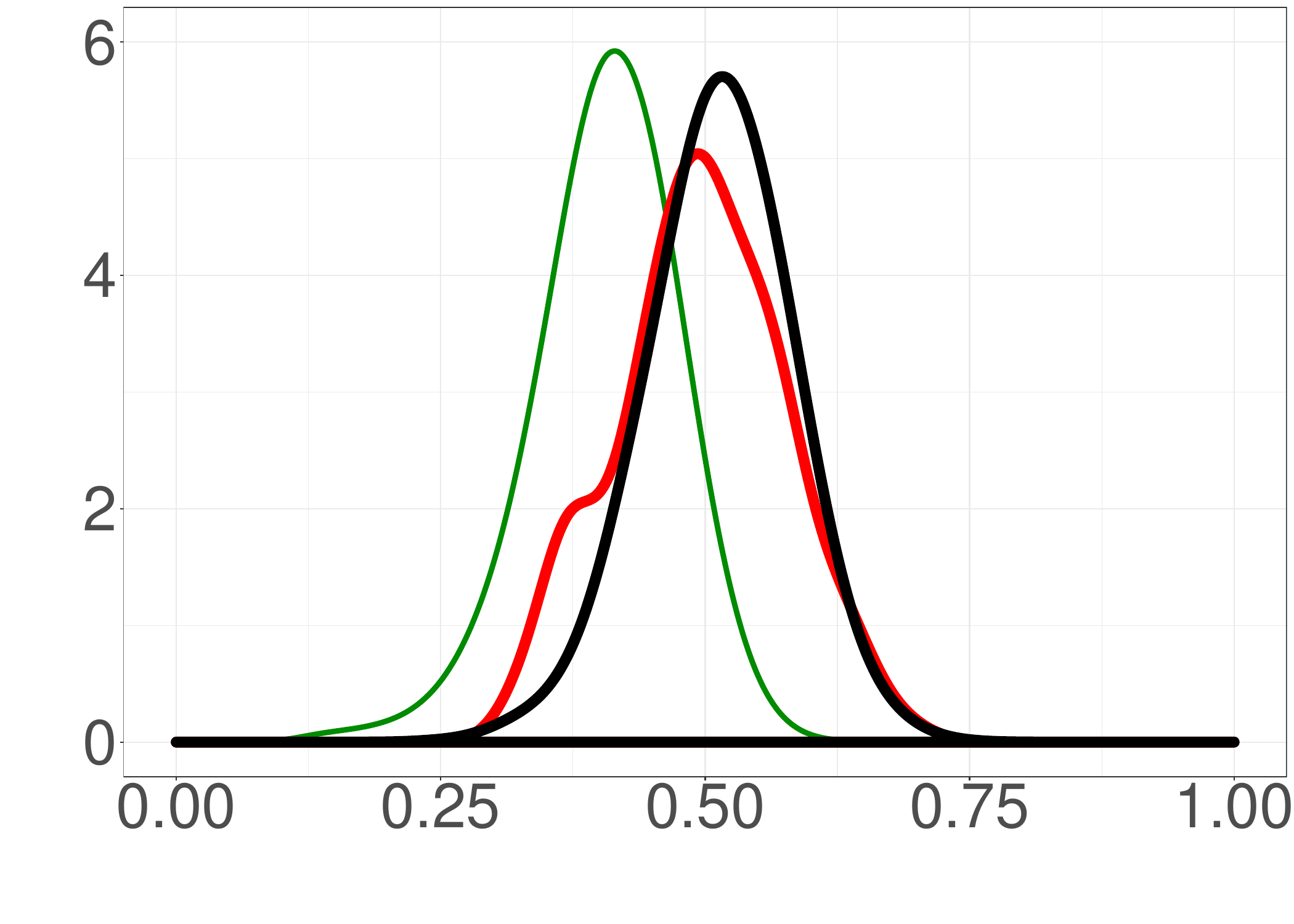}
      \label{subfig_sim_exp_mu}
  \end{subfigure}
  \begin{subfigure}{.3\textwidth}
      \centering
      \includegraphics[width=.99\linewidth]{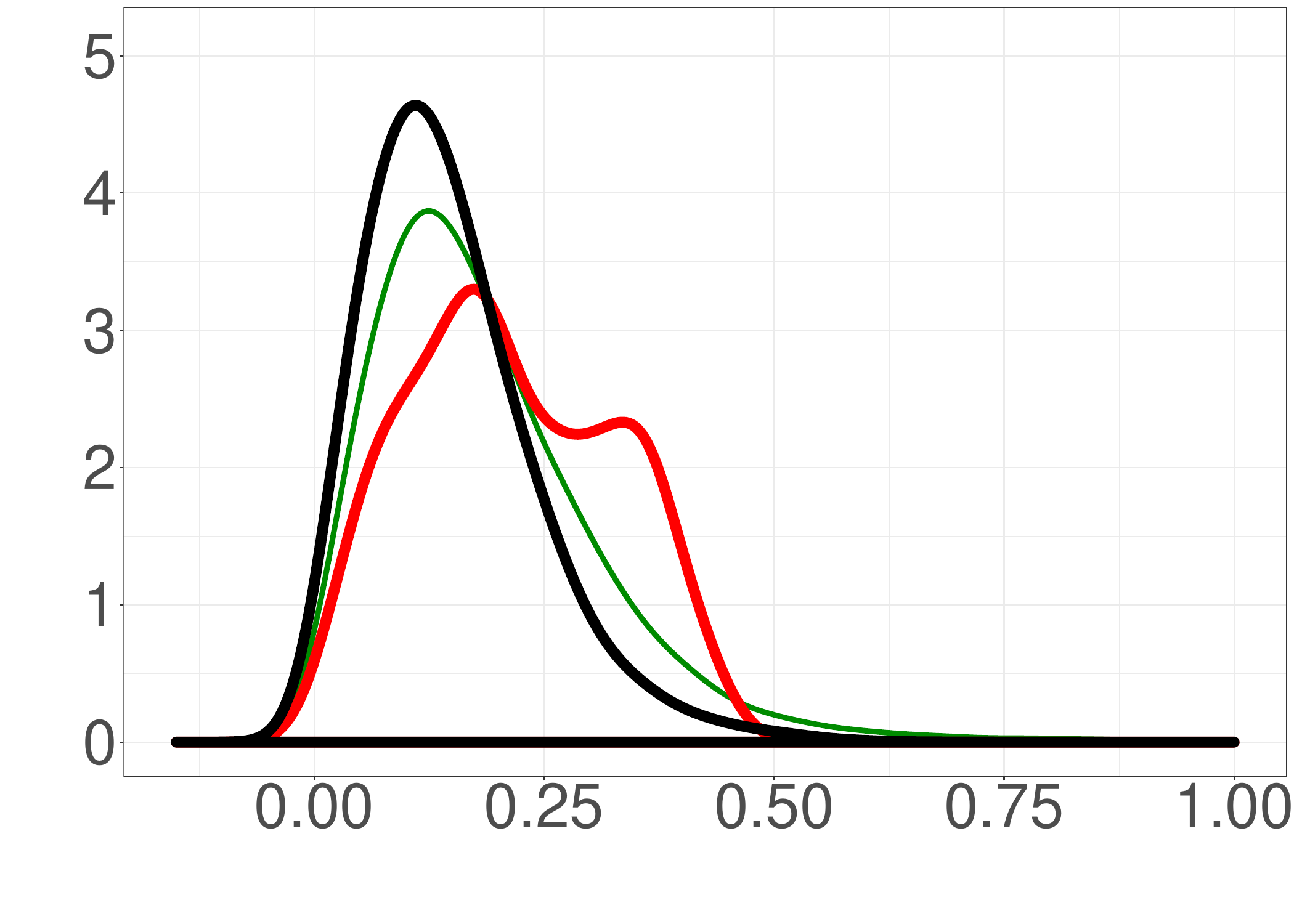}
      \label{subfig_sim_exp_K}
  \end{subfigure}
  \begin{subfigure}{.3\textwidth}
      \centering
      \includegraphics[width=.99\linewidth]{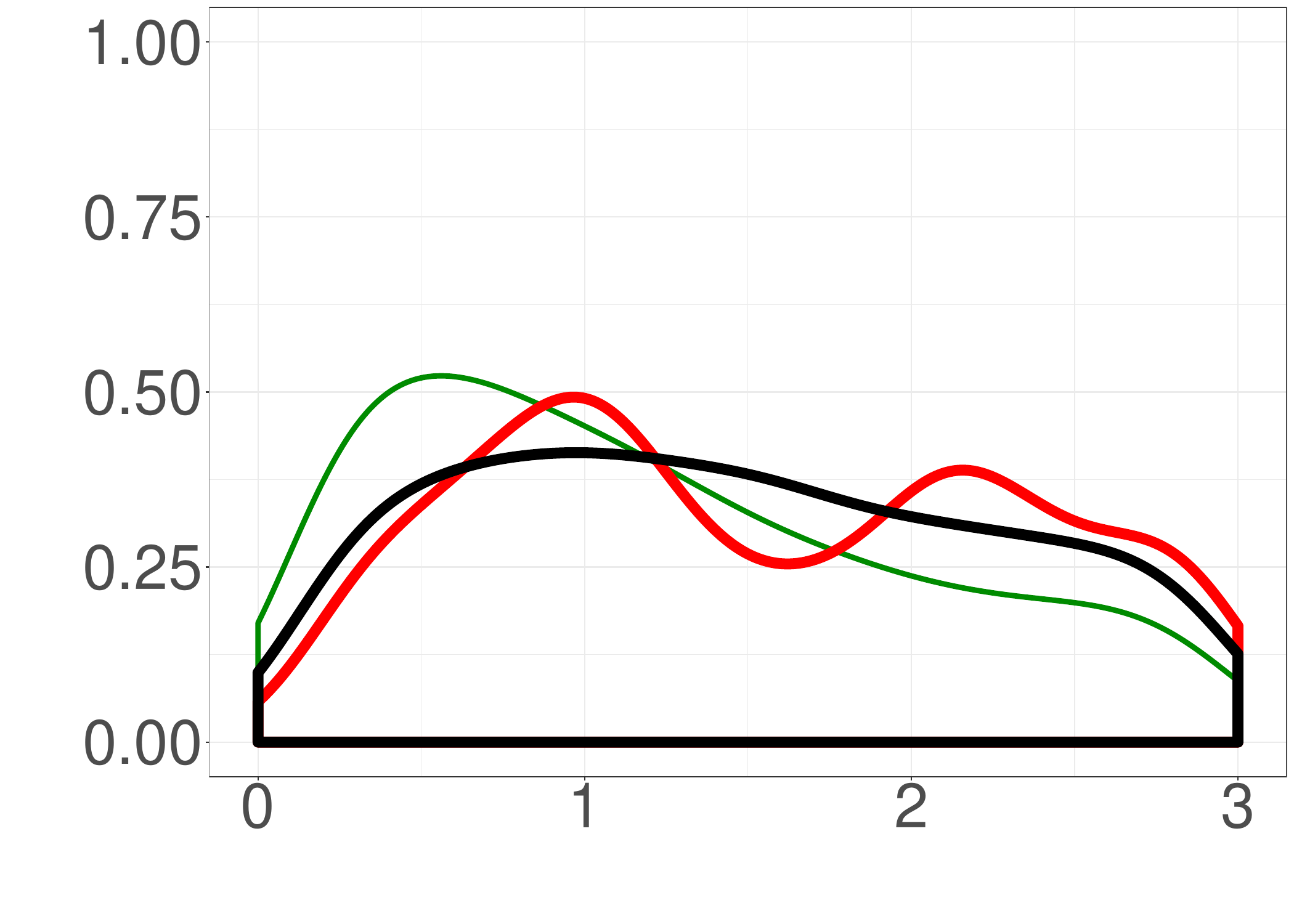}
      \label{subfig_sim_exp_beta}
  \end{subfigure}
  \begin{subfigure}{.3\textwidth}
      \centering
      \includegraphics[width=.99\linewidth]{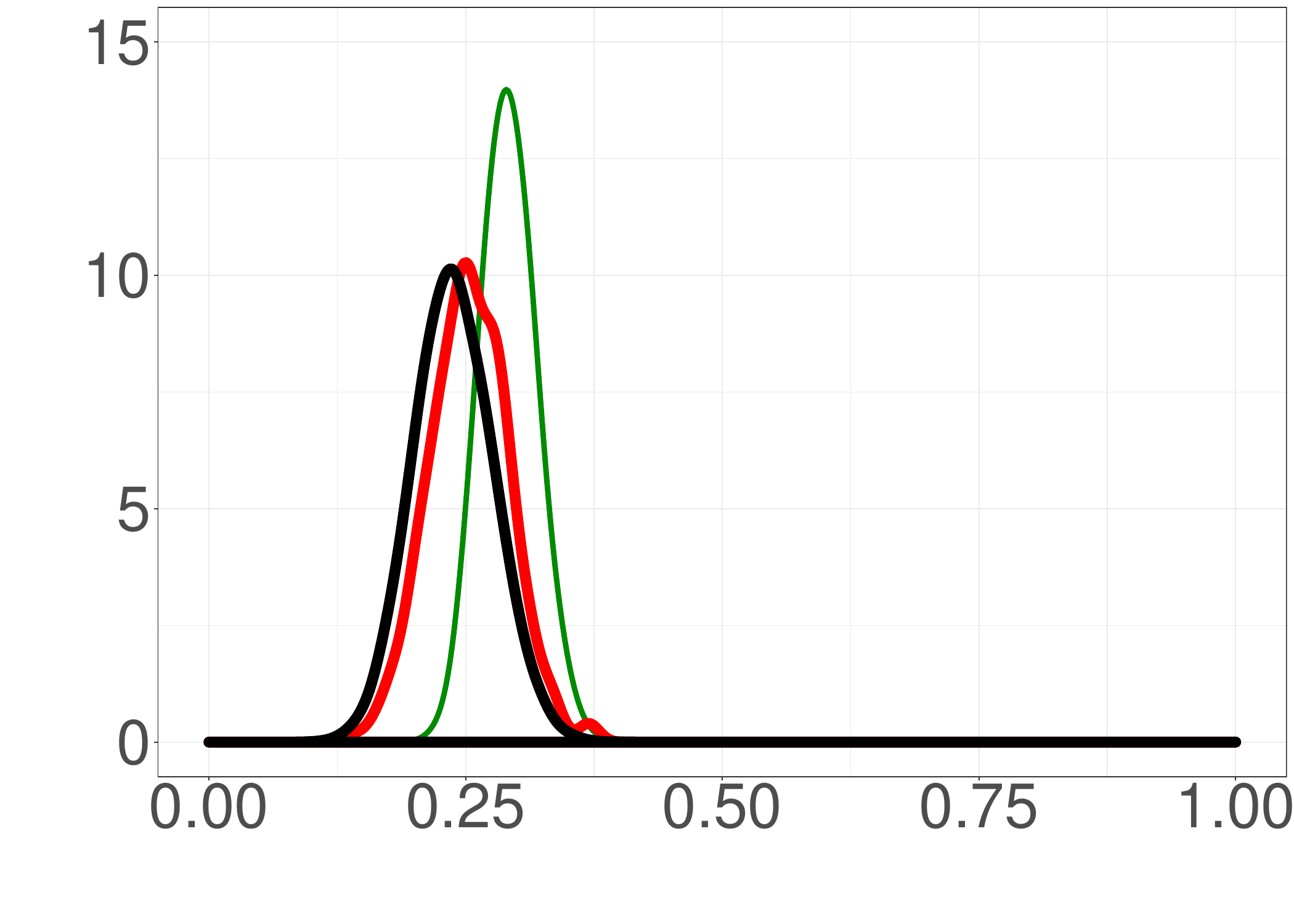}
      \caption{$\mu$}
      \label{subfig_sim_noise_mu}
  \end{subfigure}
  \begin{subfigure}{.3\textwidth}
      \centering
      \includegraphics[width=.99\linewidth]{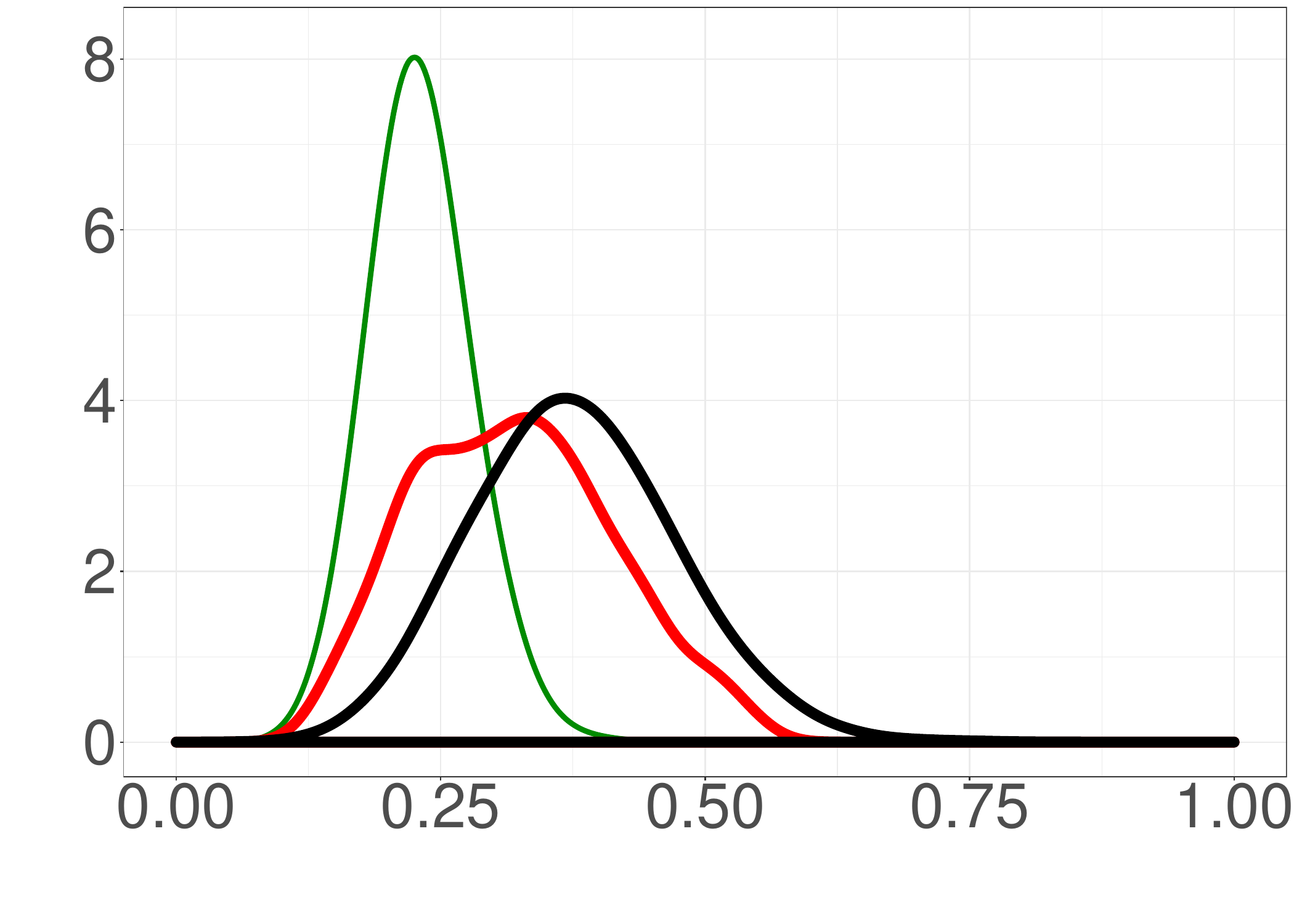}
      \caption{$K$}
      \label{subfig_sim_noise_K}
  \end{subfigure}
  \begin{subfigure}{.3\textwidth}
      \centering
      \includegraphics[width=.99\linewidth]{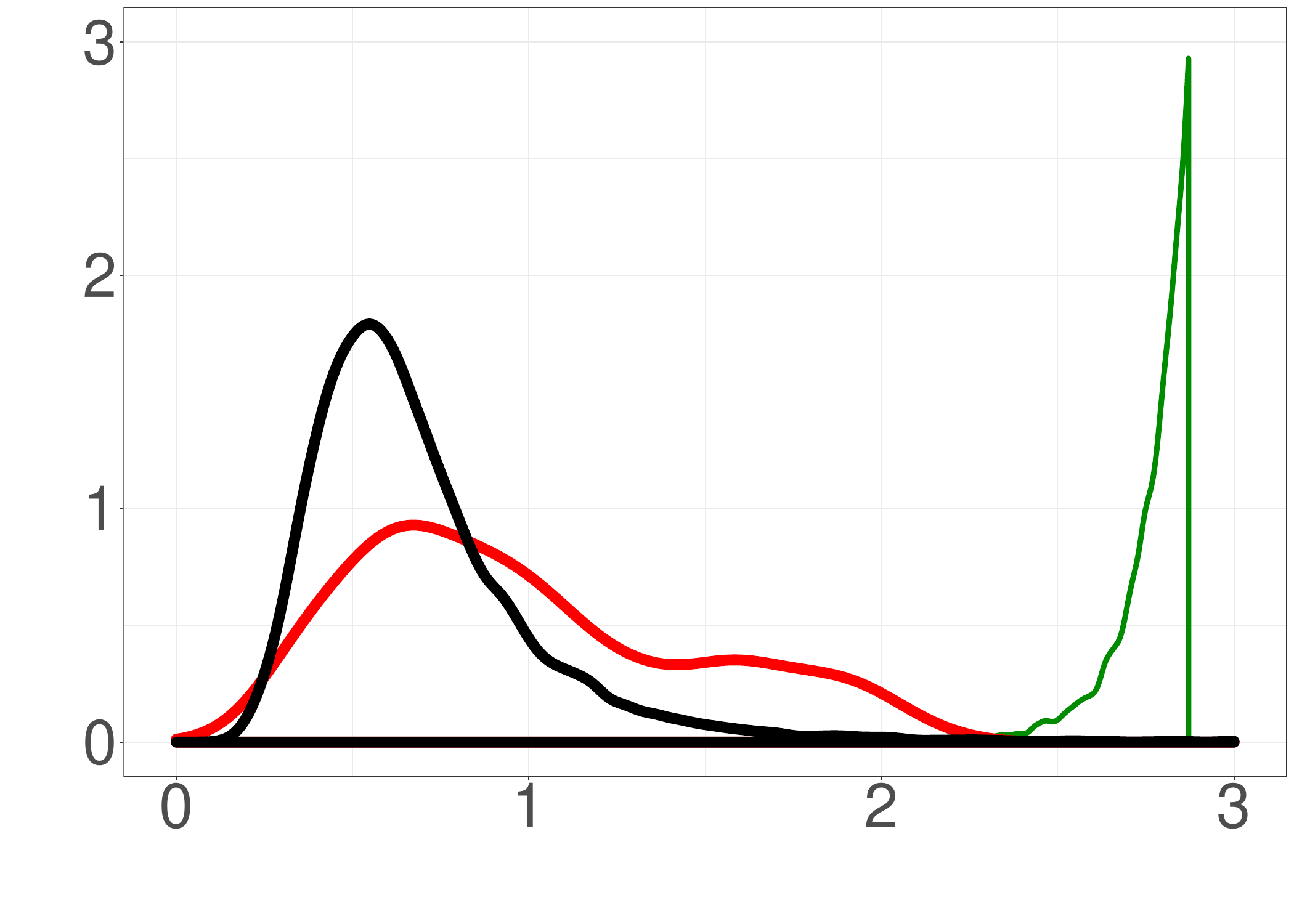}
      \caption{$\beta$}
      \label{subfig_sim_noise_beta}
  \end{subfigure}
  \caption{Distorted data posterior distributions. \textbf{Top row:} data distortion through a detection function using exponential decay. \textbf{Bottom row:} data distortion by adding noise. \textbf{Colours:}  Black represents the true posterior using the undistorted data, red represents ABC-Hawkes using the distorted data, and green represents the naive model using the distorted data. }
  \label{fig_post_sim}
  
\end{figure}

For both distorted data sets, we compute the posterior using ABC-Hawkes on the observed (distorted) data using the same Metropolis-Hastings transition kernel as above, and compare it to the idealized posterior computed on the true data. Unlike in the above missing data case, we are not aware of any other published algorithms which can handle these two types of distortion, so the only other comparison we make is to the naive method which learns the posterior using the observed data without taking the distortion into account. In Figure~\ref{fig_post_sim}, the posterior density estimates are plotted, and Table~\ref{tab_simulated} shows the posterior means and standard deviations for both scenarios. Again, ABC-Hawkes manages to learn the model parameters accurately and produces a posterior distribution which is remarkably close to the true posterior. In contrast, the naive approach is severely biased and does not get close to the true posterior. Again we note that the slight overestimation of the posterior variance is an inherent issue with ABC that comes from a necessary non-zero choice of $\epsilon_p$ \citep{li_convergence_2017}.

\begin{table}[!th]
  \caption{Simulated data posterior mean (and standard deviation)}
  \label{tab_simulated}
  \centering
  \begin{tabular}{llccc}
\toprule
Distortion &  Model   &  $\mu$ & $K$ & $\beta$ \\ 
  \midrule
   &  \textbf{True Posterior}  & \textbf{0.51 (0.07)} & \textbf{0.15 (0.09)} & \textbf{1.45 (0.80)}  \\
Lin. Deletion &  ABC-Hawkes & 0.50 (0.08) & 0.21 (0.11) & 1.54 (0.81)  \\ 
& Naive  & 0.41 (0.07) & 0.19 (0.12) & 1.22 (0.79) \\ 
 \midrule
 &  \textbf{True Posterior}  & \textbf{0.24 (0.04)} & \textbf{0.38 (0.10)} & \textbf{0.69 (0.31)}  \\
   % \cmidrule(l){2-5}
Noise   & ABC-Hawkes  & 0.25 (0.04) & 0.32 (0.09) & 0.98 (0.48) \\  
 & Naive &  0.29 (0.03) & 0.23 (0.04) & 2.89 (0.11) \\ 
  
\bottomrule
\end{tabular}
\end{table}

\section{Discussion} \label{sec_summary}

In this paper we have demonstrated that it is possible to successfully learn the parameters of a Hawkes process even when the data is distorted through mechanisms such as missingness or noise. We have based our algorithm on ABC-MCMC, which has been adapted to the unique structure of a self-exciting point process. Unlike a naive MCMC approach which ignores the potential distortion, the resulting ABC-Hawkes algorithm can learn the true posterior distribution that would have been obtained given access to the undistorted data. The strong performance of ABC-Hawkes was demonstrated using a variety of realistic data distorting mechanisms. Future research could expand the theory to other data-distorting mechanisms, for example additive noise in a multivariate setting \citep{trouleau_learning_2019} or censoring \citep{xu_learning_2017}. 

%In ABC-Hawkes, all parameters are updated at once. This could be altered according to \citet{clarte_component-wise_2020} or \citet{turner_hierarchical_2014} to component-wise updates, since the Hawkes process provides us with a notion on which summary statistics are meaningful (albeit not sufficient) for each parameter. This could potentially lead to a higher acceptance and faster convergence of the algorithm. This could potentially also be improved through sequential Monte Carlo \citep{del_moral_adaptive_2012}. 

While our simulation study focused on a simple Hawkes process with a parametrically specified self-excitation kernel, our approach is much more general than this and can be applied to other specifications of the Hawkes process as long as the resulting model is generative so that data sets $p(Y|\theta)$ can be simulated conditional on a parameter vector $\theta$. This includes recent specifications of the Hawkes process using nonparametric estimation \citep{chen_nonparametric_2016} or
LTSM networks \citep{mei_neural_2017}, which is a potential avenue for future research.

\bibliographystyle{apalike}

\bibliography{bib}

\begin{thebibliography}{}

\bibitem[\protect\citeauthoryear{Campbell and Austin}{Campbell and
  Austin}{2002}]{Campbell02}
Campbell, J.~I. and S.~Austin (2002).
\newblock Effects of response time deadlines on adults' strategy choices for
  simple addition.
\newblock {\em Memory \& Cognition\/}~{\em 30\/}(6), 988--994.

\bibitem[\protect\citeauthoryear{Chi, Feltovich, and Glaser}{Chi
  et~al.}{1981}]{Chi81}
Chi, M.~T., P.~J. Feltovich, and R.~Glaser (1981).
\newblock Categorization and representation of physics problems by experts and
  novices.
\newblock {\em Cognitive science\/}~{\em 5\/}(2), 121--152.

\bibitem[\protect\citeauthoryear{Schubert, Denmark, Crandall, Grome, and
  Pappas}{Schubert et~al.}{2013}]{Schubert13}
Schubert, C.~C., T.~K. Denmark, B.~Crandall, A.~Grome, and J.~Pappas (2013).
\newblock Characterizing novice-expert differences in macrocognition: an
  exploratory study of cognitive work in the emergency department.
\newblock {\em Annals of emergency medicine\/}~{\em 61\/}(1), 96--109.

\end{thebibliography}


\begin{thebibliography}{}

\bibitem[Arcangelis et~al., 2018]{arcangelis_overlap_2018}
Arcangelis, L.~d., Godano, C., and Lippiello, E. (2018).
\newblock The {Overlap} of {Aftershock} {Coda} {Waves} and {Short}-{Term}
  {Postseismic} {Forecasting}.
\newblock {\em Journal of Geophysical Research: Solid Earth},
  123(7):5661--5674.

\bibitem[Aryal and Jones, 2019]{aryal_fitting_2019}
Aryal, N.~R. and Jones, O.~D. (2019).
\newblock Fitting the {Bartlett}–{Lewis} rainfall model using {Approximate}
  {Bayesian} {Computation}.
\newblock {\em Mathematics and Computers in Simulation}.
\newblock In Press.

\bibitem[Beaumont et~al., 2002]{beaumont_approximate_2002}
Beaumont, M.~A., Zhang, W., and Balding, D.~J. (2002).
\newblock Approximate {Bayesian} {Computation} in {Population} {Genetics}.
\newblock {\em Genetics}, 162(4):2025--2035.

\bibitem[Bernton et~al., 2019]{bernton_approximate_2019}
Bernton, E., Jacob, P.~E., Gerber, M., and Robert, C.~P. (2019).
\newblock Approximate {Bayesian} computation with the {Wasserstein} distance.
\newblock {\em Journal of the Royal Statistical Society: Series B (Statistical
  Methodology)}, 81(2):235--269.

\bibitem[Blundell et~al., 2012]{NIPS2012_4834}
Blundell, C., Beck, J., and Heller, K.~A. (2012).
\newblock Modelling reciprocating relationships with hawkes processes.
\newblock In {\em Advances in Neural Information Processing Systems 25}, pages
  2600--2608. Curran Associates, Inc.

\bibitem[Brown, 1986]{brown_fundamentals_1986}
Brown, L.~D. (1986).
\newblock {\em Fundamentals of {Statistical} {Exponential} {Families}: {With}
  {Applications} in {Statistical} {Decision} {Theory}}.
\newblock IMS.

\bibitem[Chen and Hall, 2016]{chen_nonparametric_2016}
Chen, F. and Hall, P. (2016).
\newblock Nonparametric {Estimation} for {Self}-{Exciting} {Point}
  {Processes}—{A} {Parsimonious} {Approach}.
\newblock {\em Journal of Computational and Graphical Statistics},
  25(1):209--224.

\bibitem[Choi and Hall, 1999]{choi_nonparametric_1999}
Choi, E. and Hall, P. (1999).
\newblock Nonparametric {Approach} to {Analysis} of {Space}-{Time} {Data} on
  {Earthquake} {Occurrences}.
\newblock {\em Journal of Computational and Graphical Statistics},
  8(4):733--748.

\bibitem[Daley and Vere-Jones, 2003]{daley_introduction_2003}
Daley, D.~J. and Vere-Jones, D. (2003).
\newblock {\em An {Introduction} to the {Theory} of {Point} {Processes}:
  {Volume} {I}: {Elementary} {Theory} and {Methods}}.
\newblock Probability and {Its} {Applications}, {An} {Introduction} to the
  {Theory} of {Point} {Processes}. Springer-Verlag, New York, NY, 2 edition.

\bibitem[Donnet et~al., 2018]{donnet_nonparametric_2018}
Donnet, S., Rivoirard, V., and Rousseau, J. (2018).
\newblock Nonparametric {Bayesian} estimation of multivariate {Hawkes}
  processes.
\newblock {\em arXiv}.
\newblock arXiv: 1802.05975.

\bibitem[Ertekin et~al., 2015]{ertekin_reactive_2015}
Ertekin, {\c S}., Rudin, C., and McCormick, T.~H. (2015).
\newblock Reactive point processes: {A} new approach to predicting power
  failures in underground electrical systems.
\newblock {\em The Annals of Applied Statistics}, 9(1):122--144.

\bibitem[Fearnhead and Prangle, 2012]{fearnhead_constructing_2012}
Fearnhead, P. and Prangle, D. (2012).
\newblock Constructing summary statistics for approximate {Bayesian}
  computation: semi-automatic approximate {Bayesian} computation.
\newblock {\em Journal of the Royal Statistical Society: Series B (Statistical
  Methodology)}, 74(3):419--474.

\bibitem[Gutmann et~al., 2018]{gutmann_likelihood-free_2018}
Gutmann, M.~U., Dutta, R., Kaski, S., and Corander, J. (2018).
\newblock Likelihood-free inference via classification.
\newblock {\em Statistics and Computing}, 28(2):411--425.

\bibitem[Hawkes, 1971]{hawkes_spectra_1971}
Hawkes, A.~G. (1971).
\newblock Spectra of some self-exciting and mutually exciting point processes.
\newblock {\em Biometrika}, 58(1):83--90.

\bibitem[Hawkes and Oakes, 1974]{hawkes_cluster_1974}
Hawkes, A.~G. and Oakes, D. (1974).
\newblock A cluster process representation of a self-exciting process.
\newblock {\em Journal of Applied Probability}, 11(3):493--503.

\bibitem[Helmstetter et~al., 2006]{helmstetter_comparison_2006}
Helmstetter, A., Kagan, Y.~Y., and Jackson, D.~D. (2006).
\newblock Comparison of {Short}-{Term} and {Time}-{Independent} {Earthquake}
  {Forecast} {Models} for {Southern} {California}.
\newblock {\em Bulletin of the Seismological Society of America},
  96(1):90--106.

\bibitem[Lai et~al., 2016]{lai_topic_2016}
Lai, E.~L., Moyer, D., Yuan, B., Fox, E., Hunter, B., Bertozzi, A.~L., and
  Brantingham, P.~J. (2016).
\newblock Topic time series analysis of microblogs.
\newblock {\em IMA Journal of Applied Mathematics}, 81(3):409--431.

\bibitem[Le, 2018]{Le}
Le, T. (2018).
\newblock A multivariate {H}awkes process with gaps in observations.
\newblock {\em IEEE Transactions on Information Theory,}, 63(3):1800--1811.

\bibitem[Li et~al., 2018]{li_learning_2018}
Li, S., Xiao, S., Zhu, S., Du, N., Xie, Y., and Song, L. (2018).
\newblock Learning {Temporal} {Point} {Processes} via {Reinforcement}
  {Learning}.
\newblock In {\em Advances in {Neural} {Information} {Processing} {Systems}
  31}, pages 10781--10791. Curran Associates, Inc.

\bibitem[Li and Fearnhead, 2017]{li_convergence_2017}
Li, W. and Fearnhead, P. (2017).
\newblock Convergence of regression-adjusted approximate bayesian computation.
\newblock {\em Biometrika}, 105:301--318.

\bibitem[Linderman et~al., 2017]{linderman_bayesian_2017}
Linderman, S.~W., Wang, Y., and Blei, D.~M. (2017).
\newblock Bayesian inference for latent {Hawkes} processes.
\newblock In {\em Advances in {Approximate} {Bayesian} {Inference} {Workshop},
  31st {Conference} on {Neural} {Information} {Processing} {Systems}}.

\bibitem[Marin et~al., 2012]{marin_approximate_2012}
Marin, J.-M., Pudlo, P., Robert, C.~P., and Ryder, R.~J. (2012).
\newblock Approximate {Bayesian} computational methods.
\newblock {\em Statistics and Computing}, 22(6):1167--1180.

\bibitem[Marjoram et~al., 2003]{marjoram_markov_2003}
Marjoram, P., Molitor, J., Plagnol, V., and Tavaré, S. (2003).
\newblock Markov chain {Monte} {Carlo} without likelihoods.
\newblock {\em Proceedings of the National Academy of Sciences},
  100(26):15324--15328.

\bibitem[Mei and Eisner, 2017]{mei_neural_2017}
Mei, H. and Eisner, J.~M. (2017).
\newblock The neural hawkes process: {A} neurally self-modulating multivariate
  point process.
\newblock In {\em Advances in {Neural} {Information} {Processing} {Systems}},
  pages 6754--6764.

\bibitem[Mei et~al., 2019]{mei_imputing_2019}
Mei, H., Qin, G., and Eisner, J. (2019).
\newblock Imputing {Missing} {Events} in {Continuous}-{Time} {Event} {Streams}.
\newblock In {\em International {Conference} on {Machine} {Learning}}, pages
  4475--4485.

\bibitem[Ogata, 1981]{ogata_lewis_1981}
Ogata, Y. (1981).
\newblock On {Lewis}' simulation method for point processes.
\newblock {\em IEEE Transactions on Information Theory}, 27(1):23--31.

\bibitem[Ogata and Katsura, 2006]{ogata_immediate_2006}
Ogata, Y. and Katsura, K. (2006).
\newblock Immediate and updated forecasting of aftershock hazard.
\newblock {\em Geophysical Research Letters}, 33(10).

\bibitem[Omi et~al., 2014]{omi_estimating_2014}
Omi, T., Ogata, Y., Hirata, Y., and Aihara, K. (2014).
\newblock Estimating the {ETAS} model from an early aftershock sequence.
\newblock {\em Geophysical Research Letters}, 41(3):850--857.

\bibitem[Pritchard et~al., 1999]{pritchard_population_1999}
Pritchard, J.~K., Seielstad, M.~T., Perez-Lezaun, A., and Feldman, M.~W.
  (1999).
\newblock Population growth of human {Y} chromosomes: a study of {Y} chromosome
  microsatellites.
\newblock {\em Molecular Biology and Evolution}, 16(12):1791--1798.

\bibitem[Pudlo et~al., 2016]{pudlo_reliable_2016}
Pudlo, P., Marin, J.-M., Estoup, A., Cornuet, J.-M., Gautier, M., and Robert,
  C.~P. (2016).
\newblock Reliable {ABC} model choice via random forests.
\newblock {\em Bioinformatics}, 32(6):859--866.

\bibitem[Rambaldi et~al., 2017]{rambaldi_role_2017}
Rambaldi, M., Bacry, E., and Lillo, F. (2017).
\newblock The role of volume in order book dynamics: a multivariate {Hawkes}
  process analysis.
\newblock {\em Quantitative Finance}, 17(7):999--1020.

\bibitem[Rasmussen, 2013]{rasmussen_bayesian_2013}
Rasmussen, J.~G. (2013).
\newblock Bayesian {Inference} for {Hawkes} {Processes}.
\newblock {\em Methodology and Computing in Applied Probability},
  15(3):623--642.

\bibitem[Ripley, 1977]{ripley_1977_modelling}
Ripley, B.~D. (1977).
\newblock Modelling spatial patterns.
\newblock {\em Journal of the Royal Statistical Society: Series B
  (Methodological)}, 39(2):172--192.

\bibitem[Rizoiu et~al., 2017]{rizoiu_tutorial_2017}
Rizoiu, M.-A., Lee, Y., Mishra, S., and Xie, L. (2017).
\newblock A {Tutorial} on {Hawkes} {Processes} for {Events} in {Social}
  {Media}.
\newblock {\em arXiv}.
\newblock arXiv: 1708.06401.

\bibitem[Shelton et~al., 2018]{Shelton}
Shelton, C., Qin, Z., and Shetty, C. (2018).
\newblock Hawkes process inference with missing data.
\newblock {\em Proceedings of the Thirty-Second AAAI Conference on Artificial
  Intelligence}.

\bibitem[Shirota and Gelfand, 2017]{shirota_approximate_2017}
Shirota, S. and Gelfand, A.~E. (2017).
\newblock Approximate {Bayesian} {Computation} and {Model} {Assessment} for
  {Repulsive} {Spatial} {Point} {Processes}.
\newblock {\em Journal of Computational and Graphical Statistics},
  26(3):646--657.

\bibitem[{Stan Development Team}, 2019]{Stan_RStan_2019}
{Stan Development Team} (2019).
\newblock {RStan}: the {R} interface to {Stan}.
\newblock R package version 2.19.2.

\bibitem[Trouleau et~al., 2019]{trouleau_learning_2019}
Trouleau, W., Etesami, J., Grossglauser, M., Kiyavash, N., and Thiran, P.
  (2019).
\newblock Learning {Hawkes} {Processes} {Under} {Synchronization} {Noise}.
\newblock In {\em International {Conference} on {Machine} {Learning}}, pages
  6325--6334.

\bibitem[Tucker et~al., 2019]{tucker_handling_2019}
Tucker, D.~J., Shand, L., and Lewis, J.~R. (2019).
\newblock Handling missing data in self-exciting point process models.
\newblock {\em Spatial Statistics}, 29:160--176.

\bibitem[Upadhyay et~al., 2018]{upadhyay_deep_2018}
Upadhyay, U., De, A., and Gomez~Rodriguez, M. (2018).
\newblock Deep {Reinforcement} {Learning} of {Marked} {Temporal} {Point}
  {Processes}.
\newblock In {\em Advances in {Neural} {Information} {Processing} {Systems}
  31}, pages 3168--3178. Curran Associates, Inc.

\bibitem[Veen and Schoenberg, 2008]{Veen_2008_estimation}
Veen, A. and Schoenberg, F.~P. (2008).
\newblock Estimation of space-time branching process models in seismology using
  an em-type algorithm.
\newblock {\em Journal of the American Statistical Association},
  103(482):614--624.

\bibitem[Vihola and Franks, 2020]{vihola_use_2020}
Vihola, M. and Franks, J. (2020).
\newblock On the use of approximate {Bayesian} computation {Markov} chain
  {Monte} {Carlo} with inflated tolerance and post-correction.
\newblock {\em Biometrika}, 107(2):381--395.

\bibitem[Xiao et~al., 2017]{xiao_wasserstein_2017}
Xiao, S., Farajtabar, M., Ye, X., Yan, J., Song, L., and Zha, H. (2017).
\newblock Wasserstein {Learning} of {Deep} {Generative} {Point} {Process}
  {Models}.
\newblock In {\em Advances in {Neural} {Information} {Processing} {Systems}
  30}, pages 3247--3257. Curran Associates, Inc.

\bibitem[Xu et~al., 2017]{xu_learning_2017}
Xu, H., Luo, D., and Zha, H. (2017).
\newblock Learning {Hawkes} processes from short doubly-censored event
  sequences.
\newblock In {\em Proceedings of the 34th {International} {Conference} on
  {Machine} {Learning} - {Volume} 70}, pages 3831--3840.

\end{thebibliography}
\end{document}